\documentclass[superscriptaddress]{revtex4}
\usepackage{amsfonts}
\usepackage{amsmath}
\usepackage{amssymb}
\usepackage{graphics}

\input{tcilatex}
\begin{document}

\author{Mario Castagnino}
\affiliation{CONICET-IAFE-IFIR-Universidad de Buenos Aires, Argentina}
\author{Sebastian Fortin}
\affiliation{CONICET-IAFE-Universidad de Buenos Aires, Argentina}
\author{Olimpia Lombardi}
\affiliation{CONICET-Universidad de Buenos Aires, Argentina}

\begin{abstract}
In this paper we propose a closed-system perspective to study decoherence.
>From this perspective we analyze the spin-bath model as presented in the
literature, and a natural generalization of that model. On the basis of the
results obtained from that analysis, we argue that decoherence may be
understood as a phenomenon relative to the partition of a closed system,
selected in each particular case. This viewpoint frees the decoherence
program from certain conceptual difficulties derived from its open-system
perspective. We also argue that the usual picture of decoherence in terms of
energy dissipation is misguided.
\end{abstract}

\title{Decoherence as a relative phenomenon: a generalization of the
spin-bath model}
\maketitle

\section{Introduction}

In the literature on quantum mechanics, `decoherence'\ refers to the quantum
process that turns a coherent pure state into a decohered mixed state, which
is diagonal in a well defined basis. The phenomenon of decoherence is
essential in the account of the emergence of classicality from quantum
behavior, since it explains how interference vanishes in an extremely short
decoherence time.

The orthodox explanation of the phenomenon is given by the so-called \textit{%
environment-induced decoherence} (EID) approach (\cite{Zurek-1982}, \cite%
{Zurek-1993}, \cite{Paz-Zurek}, \cite{Zurek-2003}), according to which
decoherence is a process resulting from the interaction of an open quantum
system and its environment. In particular, by studying different physical
models, it is proved that the reduced state $\rho _{S}(t)=Tr_{E}\rho
_{SE}(t) $ of the open system rapidly diagonalizes in a well defined pointer
basis, defined case by case but not in general, which identifies the
candidates for classical states. The EID approach has been extensively
applied to many areas of physics, such as atomic physics, quantum optics and
condensed matter. In particular, the study of decoherence has acquired a
great importance in quantum computation, where the phenomenon of decoherence
represents a major obstacle to the implementation of information processing
hardware that takes advantage of superpositions. In spite of its impressive
practical success, from a conceptual viewpoint the EID approach still faces
a difficulty derived from its open-system perspective: the problem of
defining the system that decoheres.

>From the einselection view, the split of the Universe into the degrees of
freedom that are of direct interest to the observer $-$the system$-$ and the
remaining degrees of freedom $-$the environment$-$ is absolutely essential
for decoherence. \ However, since the environment may be external (a
\textquotedblleft bath\textquotedblright\ of particles interacting with the
system of interest), internal (such as collections of phonons or other
internal excitations) or a combination of both cases, the EID approach
offers no general criterion for deciding where to place the
\textquotedblleft cut\textquotedblright\ between system and environment. \
In many cases, the lack of such a general criterion leads to the need of
assuming in advance the observables that will behave classically. For
instance, in cosmology the usual strategy consists in splitting the Universe
into some degrees of freedom representing the \textquotedblleft
system\textquotedblright\ of interest, and the remaining degrees of freedom
that are supposed to be non accessible and, therefore, play the role of an
internal environment. In quantum field theory, when it is known that the
background field follows a simple classical behavior, the scalar field is
decomposed according to $\phi =\phi _{c}+\phi _{q}$, where the background
field $\phi _{c}$ plays the role of the system and the fluctuation field $%
\phi _{q}$ plays the role of the environment (see \cite{Calzetta}). Zurek
concedes that this absence of a general criterion to discriminate between
system and environment is a serious difficulty of his proposal:
\textquotedblleft \textit{In particular, one issue which has been often
taken for granted is looming big, as a foundation of the whole decoherence
program. It is the question of what are the \textquotedblleft
systems\textquotedblright\ which play such a crucial role in all the
discussions of the emergent classicality. This issue was raised earlier, but
the progress to date has been slow at best}\textquotedblright\ (see \cite%
{Zurek-1998}, p.122; for a discussion of this point, see \cite{CO-Studies}).

The main purpose of this paper is to argue that this \textquotedblleft
looming big\textquotedblright\ problem is actually a pseudo-problem, which
is simply dissolved by the fact that the split of a closed quantum system
into an open subsystem and its environment is just a way of selecting a
particular space of relevant observables of the whole closed system. But
there are many different spaces of relevant observables depending on the
observational viewpoint adopted. Therefore, the same closed system can be
decomposed in many different ways: each decomposition represents a decision
about which degrees of freedom are relevant and which can be disregarded in
each case. Since there is no privileged or \textquotedblleft
essential\textquotedblright\ decomposition, there is no need of an
unequivocal criterion for deciding where to place the cut between
\textquotedblleft the\textquotedblright\ open system and \textquotedblleft
the\textquotedblright\ environment.\ On this basis, we will show that the
usual picture of decoherence in terms of energy dissipation from the\ open
system to the\ environment can no longer be sustained. Summing up,
decoherence is a phenomenon \textit{relative} to the relevant observables
selected in each particular case. The only essential physical fact is that,
among all the observational viewpoints that may be adopted to study a
quantum system, some of them determine subspaces of relevant observables for
which the system decoheres.

For the purpose of this argumentation, the paper is organized as follows. In
Section II, by means of the concept of tensor product space, it will be
shown how the split of a whole closed system into an open system and its
environment can be understood as the selection of a space of relevant
observables. In Sections III, IV and V, the well-known spin-bath model
studied by the EID approach is presented from the perspective of the
previous subsection, and physically relevant numerical simulations are
obtained. In Sections VI, VII and VIII a generalization of the spin-bath
model is presented and solved by computer simulations; this task will allow
us to compare the results obtained for different ways of splitting the
entire closed system into an open system and its environment.\ In Section
IX, the results obtained in the previous sections are discussed from a
conceptual viewpoint in order to argue for the relative nature of
decoherence $-$which dissolves Zurek's \textquotedblleft looming
big\textquotedblright\ problem$-$ and for the rejection of the usual
description of decoherence in terms of energy dissipation. Finally, in
Section X we draw our conclusions.

\section{Selecting the relevant observables}

As it is well-known in the discussions about irreversibility, when a $-$%
classical or quantum$-$ state evolves unitarily, it cannot follow an
irreversible evolution. Therefore, if a non-unitary evolution is to be
accounted for, a further element has to be added, precisely, the split of
the maximal information about the system into a relevant part and an
irrelevant part: whereas the irrelevant part is discarded, the relevant part
may evolve non-unitarily. This idea can be rephrased in operators language.
Since the maximal information about the system is given by the space $%
\mathcal{O}$ of all its possible observables, then we restrict that maximal
information to a relevant part by selecting a subspace $\mathcal{O}%
_{R}\subset \mathcal{O}$ of relevant observables. The irreversible evolution
is the non-unitary evolution viewed from the perspective of those relevant
observables.

As emphasized by Omn\`{e}s (\cite{Omnes-2001}, \cite{Omnes-2002}),
decoherence is a particular case of irreversible process. Then, the
selection of the subspace $\mathcal{O}_{R}\subset \mathcal{O}$ is always
required in decoherence. \ In the case of the EID approach, the selection of 
$\mathcal{O}_{R}$ amounts to the partition of the whole closed system $U$
into the open system $S$ and its environment $E$ (see \cite{CFL}). In fact,
let us consider the Hilbert space $\mathcal{H}$ of the closed system $U$, $%
\mathcal{H}=\mathcal{H}_{S}\otimes $ $\mathcal{H}_{E}$, where $\mathcal{H}%
_{S}$ is the Hilbert space of $S$ and $\mathcal{H}_{E}$ the Hilbert space of 
$E$. The corresponding von Neumann-Liouville space of $U$ is $\mathcal{L}=%
\mathcal{H\otimes H=L}_{S}\otimes $ $\mathcal{L}_{E}=\mathcal{O}$, where $%
\mathcal{L}_{S}=\mathcal{H}_{S}\otimes $ $\mathcal{H}_{S}$ and $\mathcal{L}%
_{E}=\mathcal{H}_{E}\otimes $ $\mathcal{H}_{E}$. In the EID approach, the
relevant observables are those corresponding to the open system $S$: 
\begin{equation}
O_{R}=O_{S}\otimes \mathbb{I}_{E}\in \mathcal{O}_{R}\subset \mathcal{O}
\label{2.1}
\end{equation}%
where $O_{S}\in \mathcal{L}_{S}$ and $\mathbb{I}_{E}$ is the identity
operator in $\mathcal{L}_{E}$. \ The reduced density operator $\rho _{S}(t)$
of $S$ is defined by tracing over the environmental degrees of freedom, 
\begin{equation}
\rho _{S}(t)=Tr_{E}\,\rho (t)  \label{2.2}
\end{equation}%
The EID approach studies the time-evolution of $\rho _{S}(t)$ governed by an
effective master equation; it proves that, under certain definite
conditions, $\rho _{S}(t)$ converges to a stable state $\rho _{S\ast }$: 
\begin{equation}
\rho _{S}(t)\longrightarrow \rho _{S\ast }  \label{2.3}
\end{equation}%
But we also know that the expectation value of any $O_{R}\in \mathcal{O}_{R}$
in the state $\rho (t)$ of $U$ can be computed as 
\begin{equation}
\langle O_{R}\rangle _{\rho (t)}=Tr\,\left( \rho (t)(O_{S}\otimes \mathbb{I}%
_{E})\right) =Tr\left( \rho _{S}(t)\,O_{S}\right) =\langle O_{S}\rangle
_{\rho _{S}(t)}  \label{2.4}
\end{equation}%
Therefore, the convergence of $\rho _{S}(t)$ to $\rho _{S\ast }$ implies the
convergence of the expectation values: 
\begin{equation}
\langle O_{R}\rangle _{\rho (t)}=\langle O_{S}\rangle _{\rho
_{S}(t)}\longrightarrow \langle O_{S}\rangle _{\rho _{S\ast }}=\langle
O_{R}\rangle _{\rho _{\ast }}  \label{2.5}
\end{equation}%
where $\rho _{\ast }$ is a final diagonal state of the closed system $U$,
such that $\rho _{S\ast }=Tr_{E}\,\rho _{\ast }$ (see eq. (\ref{2.2}); for
details, see \cite{CFL}). This means that, although the off-diagonal terms
of $\rho (t)$ never vanish through the unitary evolution, decoherence
obtains because it is a \textit{coarse-grained process}: the system
decoheres \textit{from the observational point of view} given by any
observable belonging to the space $\mathcal{O}_{R}$.

When viewed from this closed-system perspective, the discrimination between
system and environment turns out to be the selection of the relevant
observables. By following papers \cite{Sujeeva-1} and \cite{Sujeeva-2}, we
will use the expression `\textit{tensor product structure}'\ (TPS) to call
any factorization $\mathcal{H}=\mathcal{H}_{A}\otimes \mathcal{H}_{B}$ of a
Hilbert space $\mathcal{H}$, defined by the set of observables $\left\{
O_{Ai}\otimes \mathbb{I}_{B},\mathbb{I}_{A}\otimes O_{Bi}\right\} $, such
that the eigenbases of the sets $\left\{ O_{Ai}\right\} $ and $\left\{
O_{Bi}\right\} $ are bases of $\mathcal{H}_{A}$ and $\mathcal{H}_{B}$
respectively. If $\mathcal{H}$ corresponds to a closed system $U$, the TPS $%
\mathcal{H}=\mathcal{H}_{A}\otimes \mathcal{H}_{B}$ can be viewed as
representing the decomposition of $U$ into two open systems $S_{A}$ and $%
S_{B}$, corresponding to the Hilbert spaces $\mathcal{H}_{A}$ and $\mathcal{H%
}_{B}$ respectively. In turn, given the space $\mathcal{O}=\mathcal{H}%
\otimes \mathcal{H}$ of the observables of $U$, such a decomposition
identifies the spaces $\mathcal{O}_{A}=\mathcal{H}_{A}\otimes \mathcal{H}%
_{A} $ and $\mathcal{O}_{B}=\mathcal{H}_{B}\otimes \mathcal{H}_{B}$ of the
observables of the open systems $S_{A}$ and $S_{B}$, such that $\mathcal{O}%
_{A}\otimes \mathbb{I}_{B}\subset \mathcal{O}$ and $\mathbb{I}_{A}\otimes 
\mathcal{O}_{B}\subset \mathcal{O}$. \ Once these concepts are considered,
the selection of the space $\mathcal{O}_{R}$ of relevant observables in the
EID approach turns out to amount to the selection of a particular TPS, $%
\mathcal{H}=\mathcal{H}_{S}\otimes \mathcal{H}_{E}$, such that $\mathcal{O}%
_{R}=\mathcal{O}_{S}\otimes \mathbb{I}_{E}\subset \mathcal{O}=\mathcal{H}%
\otimes \mathcal{H}$.

In this paper we will consider the particular case where the closed system $%
U $ is composed of $n$ spin-1/2 particles $P_{i}$, each represented in its
Hilbert space $\mathcal{H}_{i}$:%
\begin{equation}
\mathcal{H}=\mathcal{H}_{1}\otimes \mathcal{H}_{2}\otimes ...\otimes 
\mathcal{H}_{n}=\bigotimes\limits_{i=1}^{n}\mathcal{H}_{i}  \label{2.6}
\end{equation}%
It is quite clear that the system $U$ can be decomposed into two subsystems $%
S$ and $E$ in different ways, depending on which particles are considered as
the open system $S$. For instance, if the particle $P_{1}$ is the open
system $S$, the corresponding TPS reads%
\begin{equation}
\mathcal{H}=\mathcal{H}_{S}\otimes \mathcal{H}_{E}=\left( \mathcal{H}%
_{1}\right) \otimes \left( \bigotimes\limits_{i=2}^{n}\mathcal{H}_{i}\right)
\label{2.7}
\end{equation}%
In turn, if the particle $P_{k}$, with $1<k<n$, is viewed as the system $S$,
the corresponding TPS is%
\begin{equation}
\mathcal{H}=\mathcal{H}_{S}\otimes \mathcal{H}_{E}=\left( \mathcal{H}%
_{k}\right) \otimes \left( \bigotimes\limits_{\substack{ i=1  \\ i\neq k}}%
^{n}\mathcal{H}_{i}\right)  \label{2.8}
\end{equation}%
But we can also define the system $S$ as composed of more than a single
particle; for instance, if the particles $P_{j}$, with $j=1$ to $m<n$, are
the system $S$, the TPS in this case reads

\begin{equation}
\mathcal{H}=\mathcal{H}_{S}\otimes \mathcal{H}_{E}=\left(
\bigotimes\limits_{j=1}^{m}\mathcal{H}_{j}\right) \otimes \left(
\bigotimes\limits_{i=m+1}^{n}\mathcal{H}_{i}\right)  \label{2.9}
\end{equation}%
In the following sections we will study the phenomenon of decoherence for
different partitions of the whole closed system $U$.

\section{The spin-bath model}

\subsection{Presentation of the model}

The spin-bath model is a very simple model that has been exactly solved in
previous papers (see \cite{Zurek-1982}). Here we will study it from the
closed-system perspective presented in the previous section.

Let us consider a closed system $U=P\cup P_{1}\cup \ldots \cup P_{N}=P\cup
(\cup _{i=1}^{N}P_{i})$, where (i) $P$ is a spin-1/2 particle represented in
the Hilbert space $\mathcal{H}_{P}$, and (ii) each $P_{i}$ is a spin-1/2
particle represented in its Hilbert space $\mathcal{H}_{i}$. The Hilbert
space of the composite system $U$\ is, then, 
\begin{equation}
\mathcal{H}=\mathcal{H}_{P}\otimes \left( \bigotimes\limits_{i=1}^{N}%
\mathcal{H}_{i}\right)  \label{3.0}
\end{equation}%
In the particle $P$, the two eigenstates of the spin operator $S_{P,%
\overrightarrow{v}}$\ in direction $\overrightarrow{v}$ are $\left\vert
\Uparrow \right\rangle ,\left\vert \Downarrow \right\rangle $:%
\begin{equation}
S_{P,\overrightarrow{v}}\left\vert \Uparrow \right\rangle =\frac{1}{2}%
\left\vert \Uparrow \right\rangle \ \ \ \ \ \ \ \ S_{P,\overrightarrow{v}%
}\left\vert \Downarrow \right\rangle =-\frac{1}{2}\left\vert \Downarrow
\right\rangle  \label{3.1}
\end{equation}%
In each particle $P_{i}$, the two eigenstates of the corresponding spin
operator $S_{i,\overrightarrow{v}}$\ in direction $\overrightarrow{v}$ are $%
\left\vert \uparrow _{i}\right\rangle ,\left\vert \downarrow
_{i}\right\rangle $:%
\begin{equation}
S_{i,\overrightarrow{v}}\left\vert \uparrow _{i}\right\rangle =\frac{1}{2}%
\left\vert \uparrow _{i}\right\rangle \text{ \ \ \ \ \ \ \ }S_{i,%
\overrightarrow{v}}\left\vert \downarrow _{i}\right\rangle =-\frac{1}{2}%
\left\vert \downarrow _{i}\right\rangle  \label{3.2}
\end{equation}%
Therefore, a pure initial state of $U$ reads%
\begin{equation}
|\psi _{0}\rangle =(a\left\vert \Uparrow \right\rangle +b\left\vert
\Downarrow \right\rangle )\otimes \left( \bigotimes_{i=1}^{N}(\alpha
_{i}|\uparrow _{i}\rangle +\beta _{i}|\downarrow _{i}\rangle )\right)
\label{3.3}
\end{equation}%
where $\left\vert a\right\vert ^{2}+\left\vert b\right\vert ^{2}=1$ and $%
\left\vert \alpha _{i}\right\vert ^{2}+\left\vert \beta _{i}\right\vert
^{2}=1$. If the self-Hamiltonians $H_{P}$ of $P$ and $H_{i}$ of $P_{i}$ are
taken to be zero, and there is no interaction among the $P_{i}$, then the
total Hamiltonian $H$ of the composite system $U$ is given by the
interaction between the particle $P$ and each particle $P_{i}$ (see \cite%
{Zurek-1982}, \cite{Max}):%
\begin{equation}
H=\frac{1}{2}\left( \left\vert \Uparrow \right\rangle \left\langle \Uparrow
\right\vert -\left\vert \Downarrow \right\rangle \left\langle \Downarrow
\right\vert \right) \otimes \sum_{i=1}^{N}\left[ g_{i}\left( \left\vert
\uparrow _{i}\right\rangle \left\langle \uparrow _{i}\right\vert -\left\vert
\downarrow _{i}\right\rangle \left\langle \downarrow _{i}\right\vert \right)
\otimes \left( \bigotimes_{j\neq i}^{N}\mathbb{I}_{j}\right) \right]
\label{3.4}
\end{equation}%
where $\mathbb{I}_{j}=\left\vert \uparrow _{j}\right\rangle \left\langle
\uparrow _{j}\right\vert +\left\vert \downarrow _{j}\right\rangle
\left\langle \downarrow _{j}\right\vert $ is the identity operator on the
subspace $\mathcal{H}_{j}$. Under the action of $H$, the state $|\psi
_{0}\rangle $ evolves into%
\begin{equation}
\left\vert \psi (t)\right\rangle =a\left\vert \Uparrow \right\rangle |%
\mathcal{E}_{\Uparrow }(t)\rangle +b\left\vert \Downarrow \right\rangle |%
\mathcal{E}_{\Downarrow }(t)\rangle  \label{3.5}
\end{equation}%
where 
\begin{equation}
\left\vert \mathcal{E}_{\Uparrow }(t)\right\rangle =\left\vert \mathcal{E}%
_{\Downarrow }(-t)\right\rangle =\bigotimes_{i=1}^{N}\left( \alpha
_{i}\,e^{-ig_{i}t/2}\,\left\vert \uparrow _{i}\right\rangle +\beta
_{i}\,e^{ig_{i}t/2}\,\left\vert \downarrow _{i}\right\rangle \right)
\label{3.6}
\end{equation}

\subsection{Computing the expectation values}

The space $\mathcal{O}$ of the observables of the composite system $U$ can
be obtained as $\mathcal{O}=\mathcal{O}_{P}\otimes (\otimes _{i=1}^{N}%
\mathcal{O}_{i})$, where $\mathcal{O}_{P}$ is the space of the observables
of the particle $P$ and $\mathcal{O}_{i}$ is the space of the observables of
the particle $P_{i}$. Then, an observable $O\in \mathcal{O}=\mathcal{H}%
\otimes \mathcal{H}$ can be expressed as%
\begin{equation}
O=O_{P}\otimes (\bigotimes_{i=1}^{N}O_{i})  \label{3.7.1}
\end{equation}%
where%
\begin{eqnarray}
O_{P} &=&s_{\Uparrow \Uparrow }\left\vert \Uparrow \right\rangle
\left\langle \Uparrow \right\vert +s_{\Uparrow \Downarrow }\left\vert
\Uparrow \right\rangle \left\langle \Downarrow \right\vert +s_{\Downarrow
\Uparrow }\left\vert \Downarrow \right\rangle \left\langle \Uparrow
\right\vert +s_{\Downarrow \Downarrow }\left\vert \Downarrow \right\rangle
\left\langle \Downarrow \right\vert \ \in \mathcal{O}_{P}  \label{3.7.2} \\
O_{i} &=&\epsilon _{\uparrow \uparrow }^{(i)}|\uparrow _{i}\rangle \langle
\uparrow _{i}|+\epsilon _{\downarrow \downarrow }^{(i)}|\downarrow
_{i}\rangle \langle \downarrow _{i}|+\epsilon _{\downarrow \uparrow
}^{(i)}|\downarrow _{i}\rangle \langle \uparrow _{i}|+\epsilon _{\uparrow
\downarrow }^{(i)}|\uparrow _{i}\rangle \langle \downarrow _{i}|\ \in 
\mathcal{O}_{i}  \label{3.7.3}
\end{eqnarray}%
Since the operators $O_{P}$ and $O_{i}$ are Hermitian, the diagonal
components $s_{\Uparrow \Uparrow }$, $s_{\Downarrow \Downarrow }$, $\epsilon
_{\uparrow \uparrow }^{(i)}$, $\epsilon _{\downarrow \downarrow }^{(i)}$ are
real numbers, and the off-diagonal components are complex numbers satisfying 
$s_{\Uparrow \Downarrow }=s_{\Downarrow \Uparrow }^{\ast }$, $\epsilon
_{\uparrow \downarrow }^{(i)}=\epsilon _{\downarrow \uparrow }^{(i)\ast }$.
Then, the expectation value of the observable $O$ in the state $\left\vert
\psi (t)\right\rangle $ of eq. (\ref{3.5}) can be computed as%
\begin{equation}
\langle O\rangle _{\psi (t)}=(|a|^{2}s_{\Uparrow \Uparrow
}+|b|^{2}s_{\Downarrow \Downarrow })\,\Gamma _{0}(t)+2\func{Re}\,[ab^{\ast
}\,s_{\Downarrow \Uparrow }\,\Gamma _{1}(t)]  \label{3.8}
\end{equation}%
where (see \cite{Max}) 
\begin{eqnarray}
\Gamma _{0}(t) &=&\prod_{i=1}^{N}\left[ |\alpha _{i}|^{2}\epsilon _{\uparrow
\uparrow }^{(i)}+|\beta _{i}|^{2}\epsilon _{\downarrow \downarrow }^{(i)}+2%
\func{Re}(\alpha _{i}{}\,\beta _{i}^{\ast }\epsilon _{\downarrow \uparrow
}^{(i)}e^{ig_{i}t})\right]  \label{3.9} \\
\Gamma _{1}(t) &=&\prod_{i=1}^{N}\left[ |\alpha _{i}|^{2}\epsilon _{\uparrow
\uparrow }^{(i)}e^{ig_{i}t}+|\beta _{i}|^{2}\epsilon _{\downarrow \downarrow
}^{(i)}e^{-ig_{i}t}+2\func{Re}(\alpha _{i}{}\,\beta _{i}^{\ast }\epsilon
_{\downarrow \uparrow }^{(i)})\right]  \label{3.10}
\end{eqnarray}%
By contrast to the usual presentations, we will study two different
decompositions of the whole closed system $U$ into a relevant part and its
environment.

\section{The spin-bath model: Decomposition 1}

\subsection{Selecting the relevant observables}

In the typical situation studied by the EID approach, the open system $S$ is
the particle $P$, and the remaining particles $P_{i}$ play the role of the
environment $E$: $S=P$ and $E=\cup _{i=1}^{N}P_{i}$. Then, the TPS for this
case is 
\begin{equation}
\mathcal{H}=\mathcal{H}_{S}\otimes \mathcal{H}_{E}=\left( \mathcal{H}%
_{P}\right) \otimes \left( \bigotimes\limits_{i=1}^{N}\mathcal{H}_{i}\right)
\label{3.10-1}
\end{equation}%
Therefore, the relevant observables $O_{R}$ of the closed system $U$ are
those corresponding to the particle $P$, and they are obtained from eqs. (%
\ref{3.7.1}), (\ref{3.7.2}) and (\ref{3.7.3}), by making $\epsilon
_{\uparrow \uparrow }^{(i)}=\epsilon _{\downarrow \downarrow }^{(i)}=1$ and $%
\epsilon _{\uparrow \downarrow }^{(i)}=0$:%
\begin{equation}
O_{R}=O_{S}\otimes \mathbb{I}_{E}=\left( \sum_{s,s^{\prime }=\Uparrow
,\Downarrow }s_{ss^{\prime }}|s\rangle \langle s^{\prime }|\right) \otimes
\left( \bigotimes_{i=1}^{N}\mathbb{I}_{i}\right)  \label{3.11}
\end{equation}%
The expectation value of these observables in the state $\left\vert \psi
(t)\right\rangle $ of eq. (\ref{3.5}) is given by%
\begin{equation}
\langle O_{R}\rangle _{\psi (t)}=|a|^{2}\,s_{\Uparrow \Uparrow
}+|b|^{2}\,s_{\Downarrow \Downarrow }+2\func{Re}[ab^{\ast }\,s_{\Downarrow
\Uparrow }\,r(t)]  \label{3.12}
\end{equation}%
where 
\begin{equation}
r(t)=\langle \mathcal{E}_{\Downarrow }(t)\rangle |\mathcal{E}_{\Uparrow
}(t)\rangle =\prod_{i=1}^{N}\left( |\alpha _{i}|^{2}\,e^{-ig_{i}t}+|\beta
_{i}|^{2}\,e^{ig_{i}t}\right)  \label{3.13}
\end{equation}%
and, then, 
\begin{equation}
|r(t)|^{2}=\prod_{i=1}^{N}(|\alpha _{i}|^{4}+|\beta _{i}|^{4}+2|\alpha
_{i}|^{2}|\beta _{i}|^{2}\cos 2g_{i}t)  \label{3.14}
\end{equation}%
This means that, in eq. (\ref{3.8}), $\Gamma _{0}(t)=1$ and $\Gamma
_{1}(t)=r(t)$.

\subsection{Computing the behavior of the relevant expectation values}

In order to know the time-behavior of the expectation value of eq. (\ref%
{3.12}), we have to compute the time-behavior of $r(t)$. If we take $%
\left\vert \alpha _{i}\right\vert ^{2}$ and $\left\vert \beta
_{i}\right\vert ^{2}$ as random numbers in the closed interval $\left[ 0,1%
\right] $, such that $|\alpha _{i}|^{2}+|\beta _{i}|^{2}=1$, then 
\begin{eqnarray}
\max_{t}(|\alpha _{i}|^{4}+|\beta _{i}|^{4}+2|\alpha _{i}|^{2}|\beta
_{i}|^{2}\cos 2g_{i}t) &=&\left( \left( |\alpha _{i}|^{2}+|\beta
_{i}|^{2}\right) ^{2}\right) =1  \notag \\
\min_{t}\left( \left\vert \alpha _{i}\right\vert ^{4}+\left\vert \beta
_{i}\right\vert ^{4}+2\left\vert \alpha _{i}\right\vert ^{2}\left\vert \beta
_{i}\right\vert ^{2}\cos \left( 2g_{i}t\right) \right) &=&\left( \left(
|\alpha _{i}|^{2}-|\beta _{i}|^{2}\right) ^{2}\right) =\left( 2\left\vert
\alpha _{i}\right\vert ^{2}-1\right) ^{2}  \label{3.15}
\end{eqnarray}%
Therefore, $(|\alpha _{i}|^{4}+|\beta _{i}|^{4}+2|\alpha _{i}|^{2}|\beta
_{i}|^{2}\cos 2g_{i}t)$ is a random number which, if $t\neq 0$, fluctuates
between $1$ and $\left( 2\left\vert \alpha _{i}\right\vert ^{2}-1\right)
^{2} $. Let us notice that, when the environment has many particles (that
is, when $N\rightarrow \infty $), the statistical value of the cases $%
\left\vert \alpha _{i}\right\vert ^{2}=1$, $\left\vert \beta _{i}\right\vert
^{2}=1$, $\left\vert \alpha _{i}\right\vert ^{2}=0$ and $\left\vert \beta
_{i}\right\vert ^{2}=0$ tends to zero. In this situation, eq. (\ref{3.14})
for $|r(t)|^{2}$ is an infinite product of numbers belonging to the open
interval $\left( 0,1\right) $; as a consequence, 
\begin{equation}
\lim_{N\rightarrow \infty }r(t)=0  \label{3.16}
\end{equation}

If we know that, for $N\rightarrow \infty $, $r(t)=0$ for any $t\neq 0$ and $%
r(0)=1$, it can be expected that, for $N$ finite, $r(t)$ will evolve in time
from $r(0)=1$ to a very small value. In order to obtain the time-behavior of 
$r(t)$, different numerical simulations have been performed, where the
random $\left\vert \alpha _{i}\right\vert ^{2}$ were obtained from a
random-number generator, and the $\left\vert \beta _{i}\right\vert ^{2}$
were computed as $\left\vert \beta _{i}\right\vert ^{2}=1-\left\vert \alpha
_{i}\right\vert ^{2}$. The value of the $g_{i}$ and the time-interval $\left[
0,t_{0}\right] $ for the computations were stipulated. The time-interval $%
\left[ 0,t_{0}\right] $ was partitioned into intervals $\Delta t=t_{0}/200$,
and the function $|r(t)|^{2}$ was computed at times $t_{k}=k\Delta t$, with $%
k=0,1,...,200$, according to eq. (\ref{3.14}).\bigskip

\textbf{Simulation (a)}: The computations were performed with $N=10^{7}$, $%
N=10^{8}$ and $N=10^{9}$. All the $g_{i}$ were taken to have the same value.
The value $g_{i}=400Hz$ was selected as a reference value on the basis of
the measurement of the coupling constant in typical models of spin
interaction (\cite{XXX}). Figures 1, 2 and 3 show the time-evolution of $%
|r(t)|^{2}$ with $g_{i}=200Hz$, $g_{i}=400Hz$ and $g_{i}=800Hz$,
respectively.

 \begin{figure}[t]

 \centerline{\scalebox{0.7}{\includegraphics{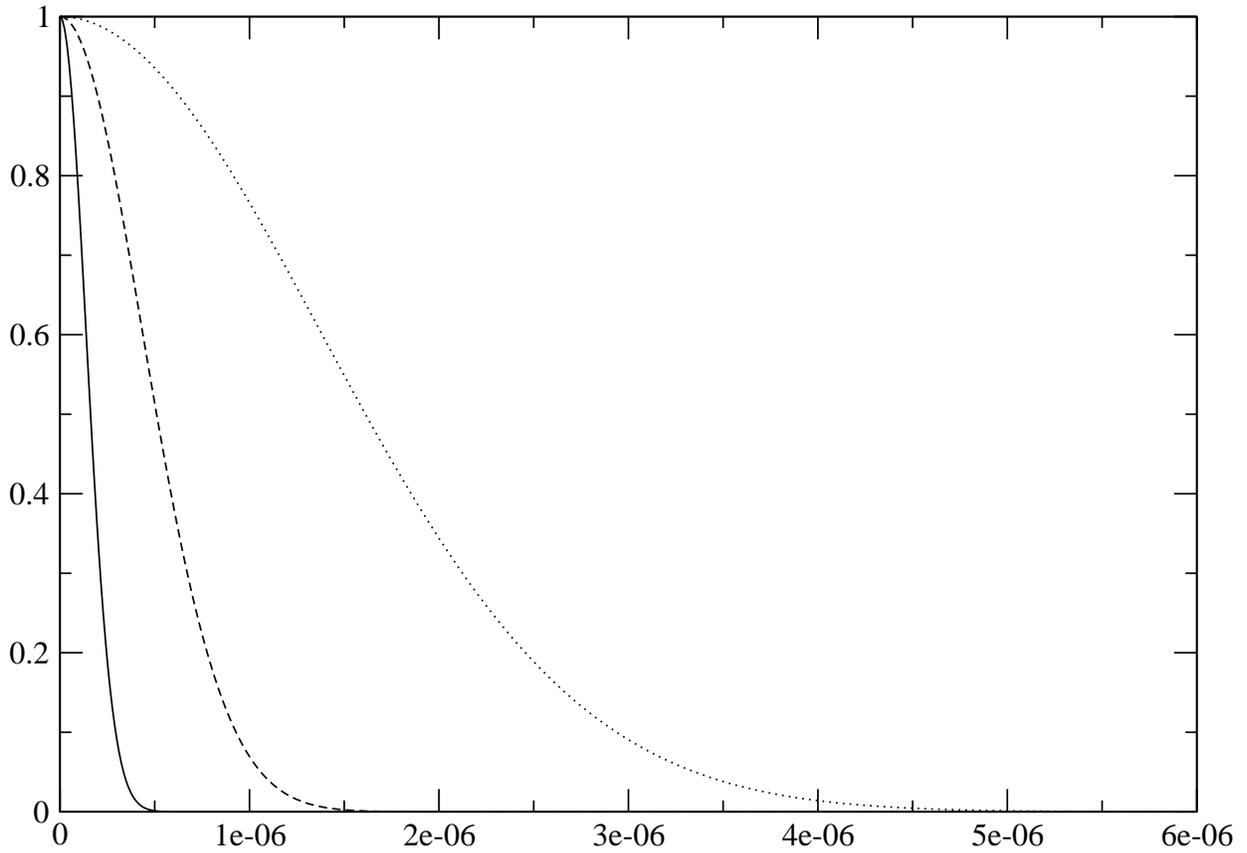}}}
\caption{Figure 1: Evolution of $%
|r(t)|^{2}$ for $g_{i}=200Hz$, and $N=10^{7}$ (dot line), $N=10^{8}$ (dash
line) and $N=10^{9}$ (solid line), with $t_{0}=6.10^{-6}s$.}
 \label{f1}\vspace*{0.cm}
\end{figure}

 \begin{figure}[t]

 \centerline{\scalebox{0.7}{\includegraphics{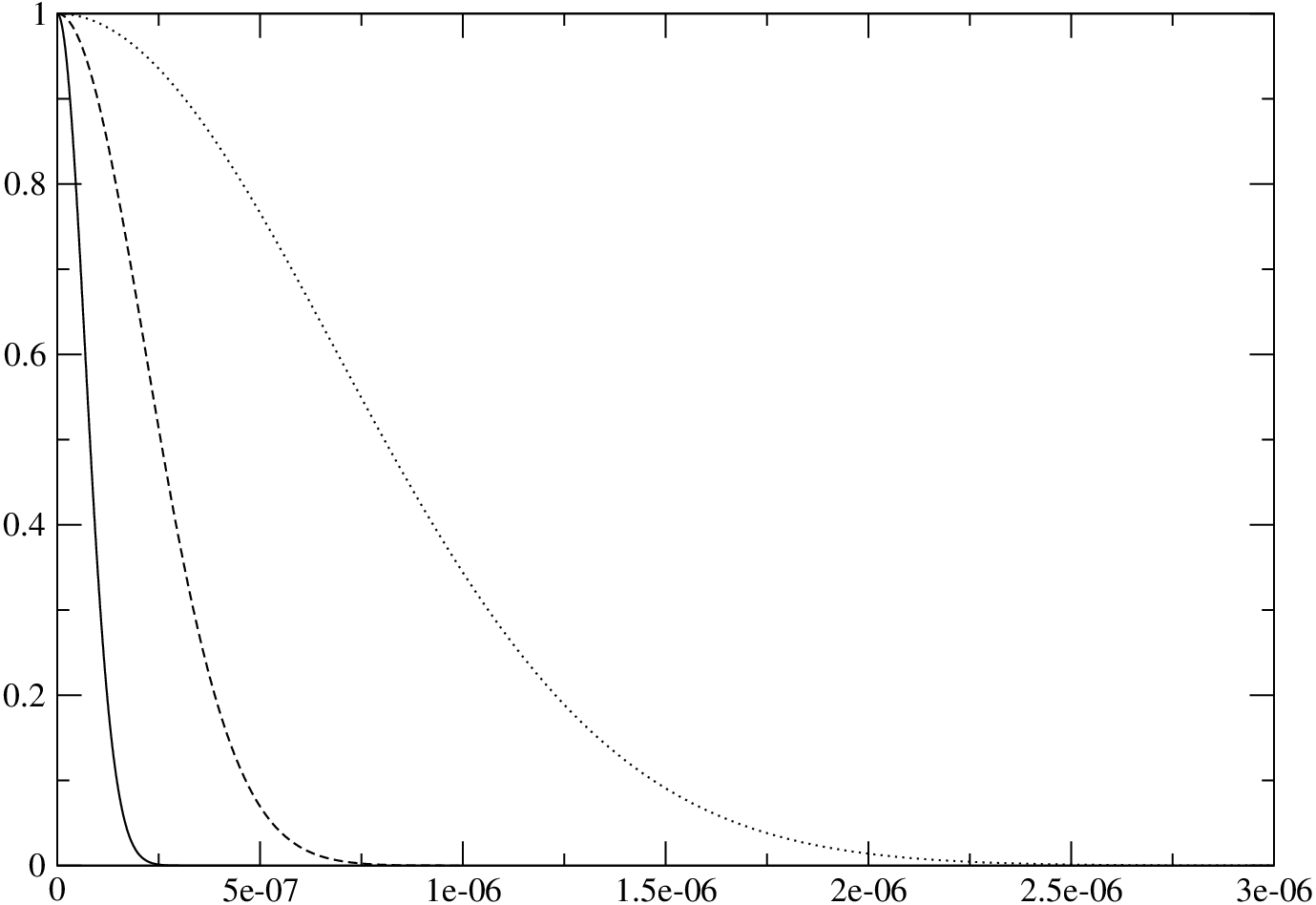}}}
\caption{Figure 2: Evolution of $%
|r(t)|^{2}$ for $g_{i}=400Hz$, and $N=10^{7}$ (dot line), $N=10^{8}$ (dash
line) and $N=10^{9}$ (solid line), with $t_{0}=3.10^{-6}s$.}
 \label{f1}\vspace*{0.cm}
\end{figure}

 \begin{figure}[t]

 \centerline{\scalebox{0.7}{\includegraphics{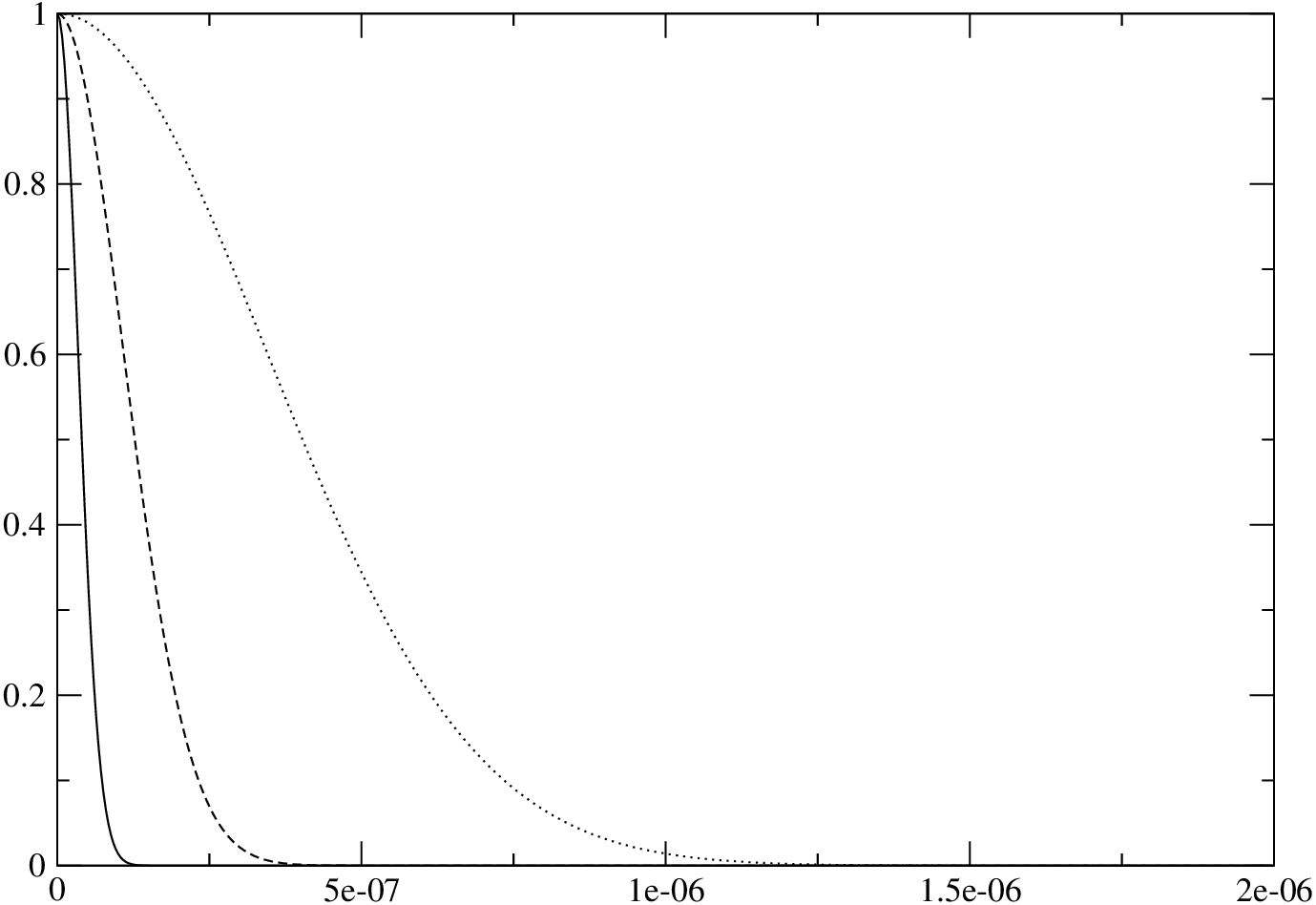}}}
\caption{Figure 3: Evolution of $%
|r(t)|^{2}$ for $g_{i}=800Hz$, and $N=10^{7}$ (dot line), $N=10^{8}$ (dash
line) and $N=10^{9}$ (solid line), with $t_{0}=2.10^{-6}s$.}
 \label{f1}\vspace*{0.cm}
\end{figure}

In these figures we can see that, as expected, (i) for any given value of $%
g_{i}$, decoherence is faster as $N$ increases, and (ii) decoherence is
faster as $g_{i}$ increases, that is, as the interaction between the
particle $P$ and each particle $P_{i}$ is stronger.\bigskip 

\textbf{Simulation (b)}: Again, the computations of $|r(t)|^{2}$ were
performed with $N=10^{7}$, $N=10^{8}$ and $N=10^{9}$, but now the values of
the $g_{i}$ were obtained from a random-number generator in the interval $%
[0,800Hz]$. The results are shown in Figure 4 for a given distribution of
the values of the $g_{i}$ (since the plots obtained for different
distributions were indistinguishable). 

 \begin{figure}[t]

 \centerline{\scalebox{0.7}{\includegraphics{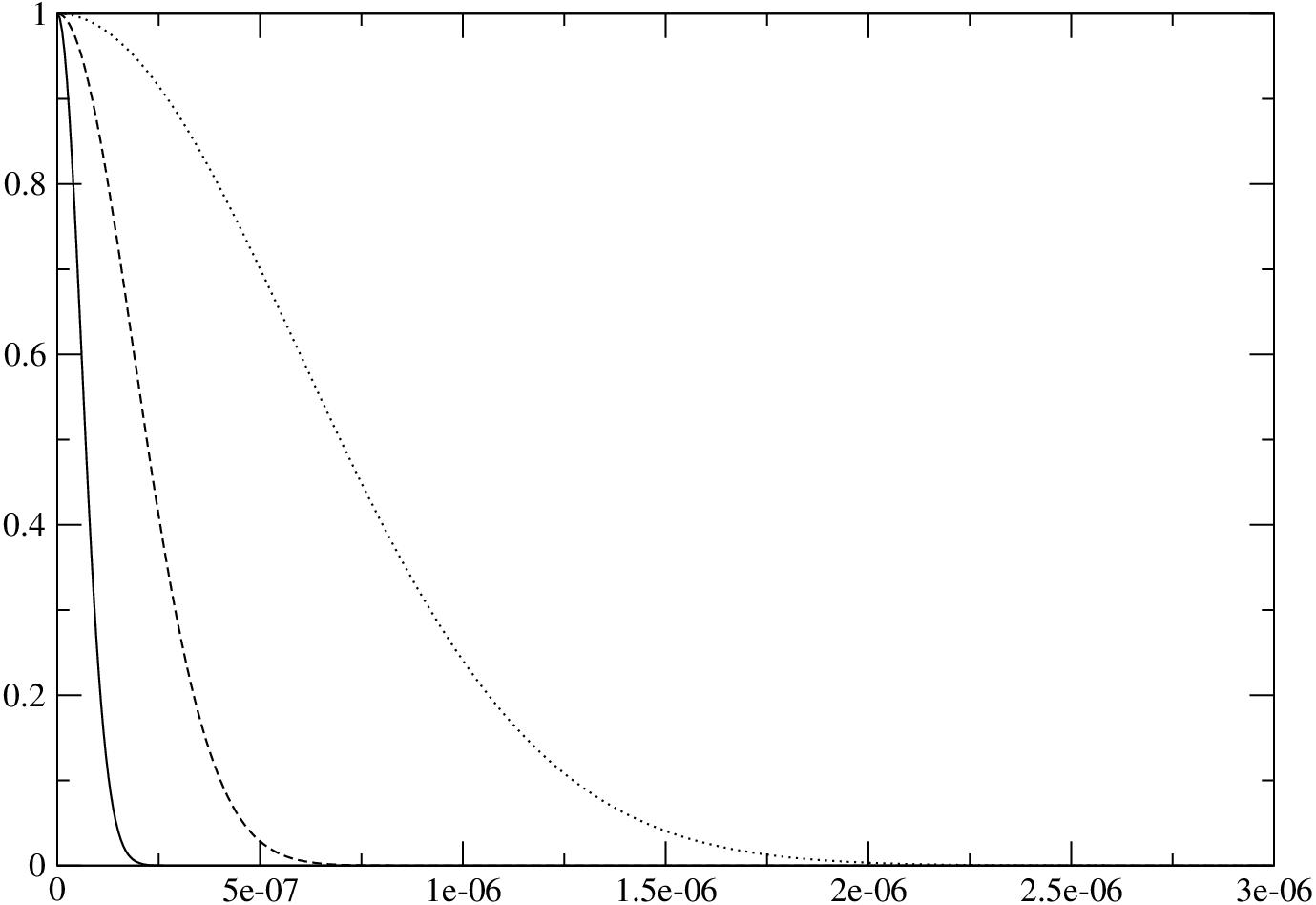}}}
\caption{Figure 4: Evolution of $%
|r(t)|^{2}$ for $g_{i}\in \left[ 0,800Hz\right] $, and $N=10^{7}$ (dot
line), $N=10^{8}$ (dash line) and $N=10^{9}$ (solid line), with $%
t_{0}=3.10^{-6}s$.}
 \label{f1}\vspace*{0.cm}
\end{figure}

Again, the figure shows that decoherence is faster as $N$ increases.
Moreover, if we compare Figure 4 ($g_{i}\in \lbrack 0,800Hz]$) with Figure 2
($g_{i}=400Hz$), we can see that the random character of the $g_{i}$
improves the \textquotedblleft efficiency\textquotedblright\ of decoherence:
the decoherence time in the case of random $g_{i}$ is shorter than in the
case of constant $g_{i}$.\bigskip 

\textbf{Simulation (c)}: In order to obtain a physically meaningful value of
the decoherence time, we performed the computation for $N=10^{20}$ (closer
to the Avogadro number) with the following strategy:

\begin{description}
\item - All the $g_{i}$ were taken to have the same value, under the
reasonable assumption that the particles of the environment are all of the
same nature and, therefore, all of them interact with the particle $P$ in
the same way. As explained above, the value of $g_{i}=400Hz$ was selected on
the basis of the typical interaction between spins.

\item - The upper limit $t_{0}$ of the time interval $\left[ 0,t_{0}\right] $
was taken as $2$ $10^{-8}s$, in order to show the physical dynamics of the
phenomenon.

\item - Of course, the running time of the computing process for $N=10^{20}$
is unattainable. Nevertheless, different runs of the computing process (with
different random values of the $\left\vert \alpha _{i}\right\vert ^{2}$ and
the $\left\vert \beta _{i}\right\vert ^{2}$) show that, for $N=10^{10}$, the
resulting plots are completely indistinguishable. Therefore, the result for $%
N=10^{20}$ can legitimately be computed by multiplying the result obtained
for $N=10^{10}$ by itself $10^{10}$ times.
\end{description}

It is interesting to see how the decoherence time decreases as $N$
increases. For this purpose, we have used the strategy of multiplying $%
|r(t)|^{2}$ corresponding to $N=10^{10}$ by itself $10^{a}$ times: $\left(
|r(t)|^{2}\right) ^{10^{a}}$ corresponds to the number of particles $%
N=10^{10+a}$. Figure 5 shows the time-evolution of $\left( |r(t)|^{2}\right)
^{10^{a}}$, corresponding to $N=10^{10+a}$, for $a=0$, $1$, $2$ and $3$.

 \begin{figure}[t]

 \centerline{\scalebox{0.7}{\includegraphics{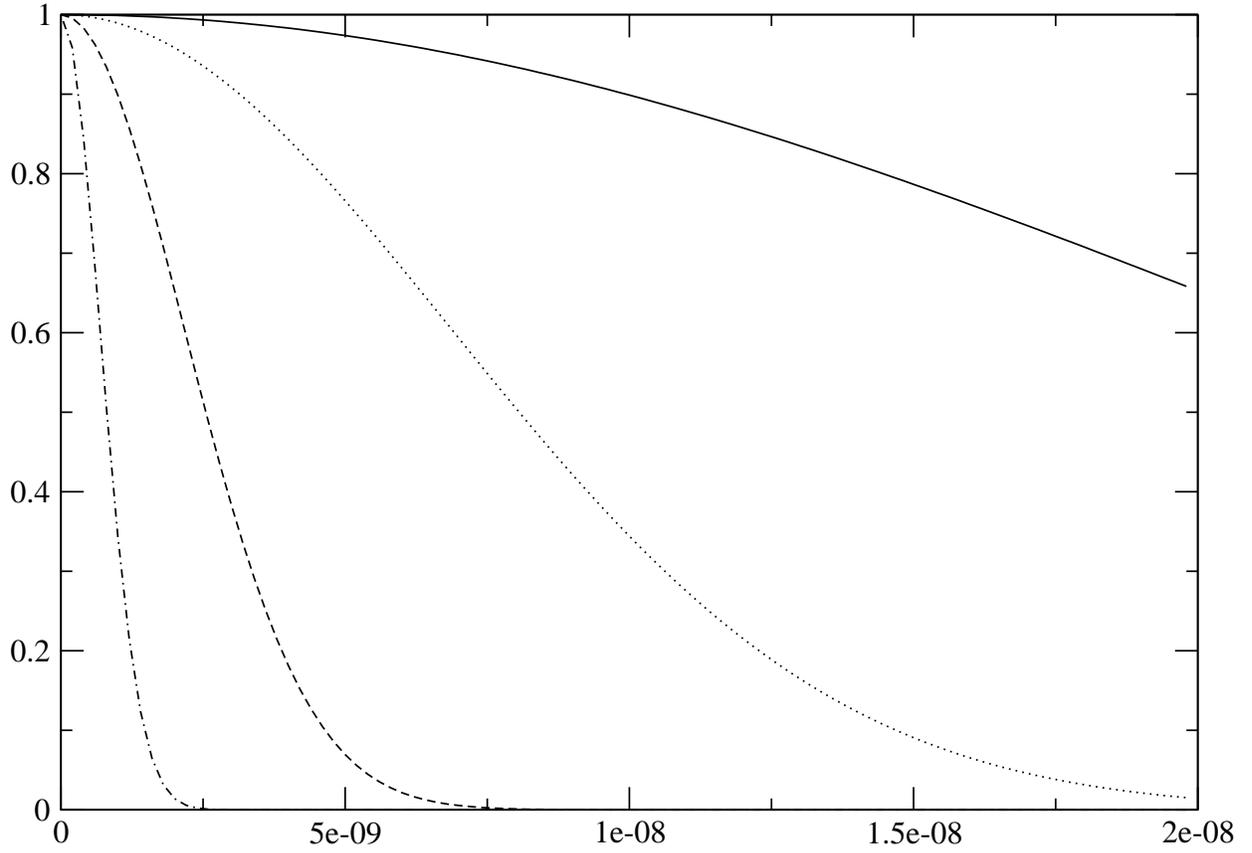}}}
\caption{Figure 5: Evolution of $%
\left( |r(t)|^{2}\right) ^{10^{a}}$ for $g_{i}=400Hz$ and $a=0$ (solid
line), $a=1$ (dot line), $a=2$\ (dash line), $a=3$ (dot-dash line), with $%
t_{0}=2.10^{-8}s$.}
 \label{f1}\vspace*{0.cm}
\end{figure}

The decoherence time was obtained by fitting the curve for $N=10^{20}$ to an
exponential, and by computing the characteristic time of the exponential.
The decoherence time so obtained was $t_{D}=10^{-13}s$, as empirically
measured (see \cite{XXX}). Let us notice\ in eq. (\ref{3.14}) that $%
|r(t)|^{2}$ comes back to its initial value when $2g_{i}t=2\pi $; then, the $%
-$recurrence$-$ Poincar\'{e} time is $t_{P}=\pi /g_{i}$. Although there is
no strict final relaxation due to the discrete nature of the model, the
relaxation time can legitimately be taken as $t_{R}\leq t_{P}/2\sim 10^{-3}s$%
. This means that, as expected, the decoherence time is many orders of
magnitude shorter than the relaxation time.

\section{The spin-bath model: Decomposition 2}

\subsection{Selecting the relevant observables}

Although in the usual presentations of the model the open system of interest
is $P$, we can conceive different ways of splitting the whole closed system $%
U$ into an open system $S$ and its environment $E$. For instance, we can
decide to observe a particular particle $P_{j}$ of what was previously
considered the environment, and to consider the remaining particles as the
new environment, in such a way that $S=P_{j}$ and $E=P\cup (\cup _{i=1,i\neq
j}^{N}P_{i})$. The total Hilbert space of the closed composite system $U$\
is still given by eq. (\ref{3.0}), but in this case the corresponding TPS is 
\begin{equation}
\mathcal{H}=\mathcal{H}_{S}\otimes \mathcal{H}_{E}=\left( \mathcal{H}%
_{j}\right) \otimes \left( \mathcal{H}_{P}\otimes \left( \bigotimes\limits 
_{\substack{ i=1  \\ i\neq j}}^{N}\mathcal{H}_{i}\right) \right)  \label{4-0}
\end{equation}%
and the relevant observables $O_{R}$ of the closed system $U$ are those
corresponding to the particle $P_{j}$:%
\begin{equation}
O_{R}=O_{S}\otimes \mathbb{I}_{E}=O_{P_{j}}\otimes \left( \mathbb{I}%
_{P}\otimes \left( \bigotimes_{\substack{ i=1  \\ i\neq j}}^{N}\mathbb{I}%
_{i}\right) \right)  \label{4-1.1}
\end{equation}%
where (see eq. (\ref{3.7.3})) 
\begin{equation}
O_{P_{j}}=\epsilon _{\uparrow \uparrow }^{(j)}\,|\uparrow _{j}\rangle
\langle \uparrow _{j}|+\epsilon _{\downarrow \downarrow }^{(j)}\,|\downarrow
_{j}\rangle \langle \downarrow _{j}|+\epsilon _{\downarrow \uparrow
}^{(j)}\,|\downarrow _{j}\rangle \langle \uparrow _{j}|+\epsilon _{\uparrow
\downarrow }^{(j)}\,|\uparrow _{j}\rangle \langle \downarrow _{j}|
\label{4-1.2}
\end{equation}%
$\mathbb{I}_{P}$ is the identity operator on the subspace $\mathcal{H}_{P}$,
and the coefficients $\epsilon _{\uparrow \uparrow }^{(j)}$, $\epsilon
_{\downarrow \downarrow }^{(j)}$, $\epsilon _{\downarrow \uparrow }^{(j)}$
are now generic. The expectation value of the observables $O_{R}$ in the
state $\left\vert \psi (t)\right\rangle $ of eq. (\ref{3.5}) is given by%
\begin{equation}
\langle O_{R}\rangle _{\psi (t)}=\langle \psi (t)|O_{R_{j}}|\psi (t)\rangle
=\left\vert \alpha _{j}\right\vert ^{2}\epsilon _{\uparrow \uparrow
}^{(j)}+\left\vert \beta _{j}\right\vert ^{2}\epsilon _{\downarrow
\downarrow }^{(j)}+2\func{Re}\left( \alpha _{j}\beta _{j}^{\ast }\epsilon
_{\downarrow \uparrow }^{(j)}e^{ig_{j}t}\right)  \label{4-1.3}
\end{equation}

\subsection{Computing the behavior of the relevant expectation values}

In order to know the time-behavior of the expectation value of the $O_{R}$,
we have to compute the time-behavior of the third term of eq. (\ref{4-1.3}),
which can be rewritten as%
\begin{equation}
2\func{Re}\left( \alpha _{j}\beta _{j}^{\ast }\epsilon _{\downarrow \uparrow
}^{(j)}e^{ig_{j}t}\right) =2\left\vert \alpha _{j}\beta _{j}^{\ast }\epsilon
_{\downarrow \uparrow }^{(j)}\right\vert \cos \left( g_{j}t+\arg \left(
\alpha _{j}\beta _{j}^{\ast }\epsilon _{\downarrow \uparrow }^{(j)}\right)
\right)  \label{4-1.4}
\end{equation}%
In this case, a numerical simulation is not necessary to see that eq. (\ref%
{4-1.4}) is an oscillating function which, as a consequence, has no limit
for $t\rightarrow \infty $. This result is not surprising, but completely
reasonable from a physical point of view. In fact, with the exception of the
particle $P$, the remaining particles of the environment $E$ are uncoupled
to each other: each $P_{i}$ evolves as a free system and, for this reason, $%
E $ is unable to reach a final stable state.

\section{A generalized spin-bath model}

\subsection{Presentation of the model}

Let us consider a closed system $U=A\cup B$ where:

\begin{enumerate}
\item[(i)] The subsystem $A$ is composed of $M$ spin-1/2 particles $A_{i}$,
with $i=1,2,...,M$, each one of them represented in its Hilbert space $%
\mathcal{H}_{A_{i}}$. In each $A_{i}$, the two eigenstates of the spin
operator $S_{A_{i},\overrightarrow{v}}$ in direction $\overrightarrow{v}$\
are $\left\vert \Uparrow _{i}\right\rangle $ and $\left\vert \Downarrow
_{i}\right\rangle $:%
\begin{equation}
S_{A_{i},\overrightarrow{v}}\left\vert \Uparrow _{i}\right\rangle =\frac{1}{2%
}\left\vert \Uparrow _{i}\right\rangle \text{ \ \ \ \ \ \ \ }S_{A_{i},%
\overrightarrow{v}}\left\vert \Downarrow _{i}\right\rangle =-\frac{1}{2}%
\left\vert \Downarrow _{i}\right\rangle  \label{4.1}
\end{equation}%
The Hilbert space of $A$ is $\mathcal{H}_{A}=\bigotimes\limits_{i=1}^{M}%
\mathcal{H}_{A_{i}}$. Then, a pure initial state of $A$ reads%
\begin{equation}
\left\vert \psi _{A}\right\rangle =\bigotimes_{i=1}^{M}\left(
a_{i}\left\vert \Uparrow _{i}\right\rangle +b_{i}\left\vert \Downarrow
_{i}\right\rangle \right) ,\ \ \text{with\ }\left\vert a_{i}\right\vert
^{2}+\left\vert b_{i}\right\vert ^{2}=1  \label{4.2}
\end{equation}

\item[(ii)] The subsystem $B$ is composed of $N$ spin-1/2 particles $B_{k}$,
with $k=1,2,...,N$, each one of them represented in its Hilbert space $%
\mathcal{H}_{B_{k}}$. In each $B_{k}$, the two eigenstates of the spin
operator $S_{B_{k},\overrightarrow{v}}$ in direction $\overrightarrow{v}$\
are $\left\vert \uparrow _{k}\right\rangle $ and $\left\vert \downarrow
_{k}\right\rangle $:%
\begin{equation}
S_{B_{k},\overrightarrow{v}}\left\vert \uparrow _{k}\right\rangle =\frac{1}{2%
}\left\vert \uparrow _{k}\right\rangle \text{ \ \ \ \ \ \ \ }S_{B_{k},%
\overrightarrow{v}}\left\vert \downarrow _{k}\right\rangle =-\frac{1}{2}%
\left\vert \downarrow _{k}\right\rangle  \label{4.3}
\end{equation}%
The Hilbert space of $B$ is $\mathcal{H}_{B}=\bigotimes\limits_{k=1}^{N}%
\mathcal{H}_{B_{k}}$. Then, a pure initial state of $B$ reads%
\begin{equation}
\left\vert \psi _{B}\right\rangle =\bigotimes_{k=1}^{N}\left( \alpha
_{k}\left\vert \uparrow _{k}\right\rangle +\beta _{k}\left\vert \downarrow
_{k}\right\rangle \right) \text{, \ \ with \ \ }\left\vert \alpha
_{k}\right\vert ^{2}+\left\vert \beta _{k}\right\vert ^{2}=1  \label{4.4}
\end{equation}
\end{enumerate}

The Hilbert space of the composite system $U=A\cup B$\ is, then, 
\begin{equation}
\mathcal{H}=\mathcal{H}_{A}\otimes \mathcal{H}_{B}=\left(
\bigotimes\limits_{i=1}^{M}\mathcal{H}_{A_{i}}\right) \otimes \left(
\bigotimes\limits_{k=1}^{N}\mathcal{H}_{B_{k}}\right)  \label{4.5}
\end{equation}%
Therefore, from eqs. (\ref{4.2}) and (\ref{4.4}), a pure initial state of $U$
reads%
\begin{equation}
\left\vert \psi _{0}\right\rangle =\left\vert \psi _{A}\right\rangle \otimes
\left\vert \psi _{B}\right\rangle =\left( \bigotimes_{i=1}^{M}\left(
a_{i}\left\vert \Uparrow _{i}\right\rangle +b_{i}\left\vert \Downarrow
_{i}\right\rangle \right) \right) \otimes \left( \bigotimes_{k=1}^{N}\left(
\alpha _{k}\left\vert \uparrow _{k}\right\rangle +\beta _{k}\left\vert
\downarrow _{k}\right\rangle \right) \right)  \label{4.6}
\end{equation}

As in the original spin-bath model, the self-Hamiltonians $H_{A_{i}}$ and $%
H_{B_{k}}$ are taken to be zero. In turn, there is no interaction among the
particles $A_{i}$ nor among the particles $B_{k}$. As a consequence, the
total Hamiltonian $H$ of the composite system $U$ is given by (see eq. (\ref%
{3.4}))%
\begin{equation}
H=H_{A}\otimes H_{B}=\left( \sum_{i=1}^{M}\left[ \frac{1}{2}\left(
\left\vert \Uparrow _{i}\right\rangle \left\langle \Uparrow _{i}\right\vert
-\left\vert \Downarrow _{i}\right\rangle \left\langle \Downarrow
_{i}\right\vert \right) \otimes \left( \bigotimes_{j\neq i}^{M}\mathbb{I}%
_{A_{j}}\right) \right] \right) \otimes \left( \sum_{k=1}^{N}\left[
g_{k}\left( \left\vert \uparrow _{k}\right\rangle \left\langle \uparrow
_{k}\right\vert -\left\vert \downarrow _{k}\right\rangle \left\langle
\downarrow _{k}\right\vert \right) \otimes \left( \bigotimes_{l\neq k}^{N}%
\mathbb{I}_{B_{l}}\right) \right] \right)   \label{4.7}
\end{equation}%
where $\mathbb{I}_{A_{j}}=\left\vert \Uparrow _{j}\right\rangle \left\langle
\Uparrow _{j}\right\vert +\left\vert \Downarrow _{j}\right\rangle
\left\langle \Downarrow _{j}\right\vert $ is the identity on the subspace $%
\mathcal{H}_{A_{j}}$ and$\ \mathbb{I}_{B_{l}}=\left\vert \uparrow
_{l}\right\rangle \left\langle \uparrow _{l}\right\vert +\left\vert
\downarrow _{l}\right\rangle \left\langle \downarrow _{l}\right\vert $ is
the identity on the subspace $\mathcal{H}_{B_{l}}$. Let us notice that the
eq. (\ref{3.4}) of the original model is the particular case of eq. (\ref%
{4.7}) for $M=1$. This Hamiltonian describes a situation where the particles
of $A$ do not interact to each other, the same holds for the particles of $B$%
, but each particle of $A$ interacts with all the particles of $B$ and vice
versa, as shown in Figure 6.%

 \begin{figure}[t]

 \centerline{\scalebox{0.7}{\includegraphics{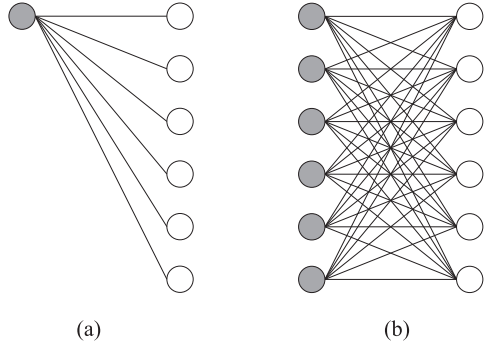}}}
\caption{Figure 6: Schema of the
interactions among the particles of the open system $A$ (grey circles) and
of the open system $B$ (white circles): (a) original spin-bath model ($M=1$%
), and (b) generalized spin-bath model ($M\neq 1$).}
 \label{f1}\vspace*{0.cm}
\end{figure}

In eq. (\ref{4.7}), $H$ is written in its diagonal form; then, the energy
eigenvectors are%
\begin{align}
& \left\vert \Uparrow _{1}\right\rangle ...\left\vert \Uparrow
_{i}\right\rangle ...\left\vert \Uparrow _{M-1}\right\rangle \left\vert
\Uparrow _{M}\right\rangle \left\vert \uparrow _{1}\right\rangle
...\left\vert \uparrow _{k}\right\rangle ...\left\vert \uparrow
_{N-1}\right\rangle \left\vert \uparrow _{N}\right\rangle   \notag \\
& \left\vert \Uparrow _{1}\right\rangle ...\left\vert \Uparrow
_{i}\right\rangle ...\left\vert \Uparrow _{M-1}\right\rangle \left\vert
\Uparrow _{M}\right\rangle \left\vert \uparrow _{1}\right\rangle
...\left\vert \uparrow _{k}\right\rangle ...\left\vert \uparrow
_{N-1}\right\rangle \left\vert \downarrow _{N}\right\rangle   \notag \\
& ...  \notag \\
& \left\vert \Downarrow _{1}\right\rangle ...\left\vert \Downarrow
_{i}\right\rangle ...\left\vert \Downarrow _{M-1}\right\rangle \left\vert
\Downarrow _{M}\right\rangle \left\vert \downarrow _{1}\right\rangle
...\left\vert \downarrow _{k}\right\rangle ...\left\vert \downarrow
_{N-1}\right\rangle \left\vert \downarrow _{N}\right\rangle   \label{4.8}
\end{align}%
In turn, the eigenvectors of $H_{A}$ form a basis of $\mathcal{H}_{A}$. In
order to simplify the expressions, we will introduce a particular
arrangement into the set of those vectors, by calling them $\left\vert 
\mathcal{A}_{i}\right\rangle $:\ the set $\left\{ \left\vert \mathcal{A}%
_{i}\right\rangle \right\} $ is an eigenbasis of $H_{A}$ with $2^{M}$
elements. The $\left\vert \mathcal{A}_{i}\right\rangle $ will be ordered in
terms of the number $l\in 
\mathbb{N}
_{0}$ of particles of $A$ having spin $\left\vert \Downarrow \right\rangle $%
. Then, we have that:

\begin{itemize}
\item $l=0$ corresponds to the unique state with all the particles with spin 
$\left\vert \Uparrow \right\rangle $:%
\begin{equation}
\left\vert \mathcal{A}_{1}\right\rangle =\left\vert \Uparrow ,\Uparrow
,...,\Uparrow ,\Uparrow \right\rangle \Longrightarrow H_{A}\left\vert 
\mathcal{A}_{1}\right\rangle =\frac{M}{2}\left\vert \mathcal{A}%
_{1}\right\rangle  \label{4.9}
\end{equation}

\item $l=1$ corresponds to the $M$ states with only one particle with spin $%
\left\vert \Downarrow \right\rangle $. Since the order of the eigenvectors
with the same eigenvalue will be irrelevant for the computations, we will
order these states in an arbitrary way:%
\begin{eqnarray}
\left\vert \mathcal{A}_{j}\right\rangle &=&\left\vert \Uparrow ,\Uparrow
,...,\Uparrow ,\Downarrow ,\Uparrow ,...,\Uparrow ,\Uparrow \right\rangle
\Longrightarrow H_{A}\left\vert \mathcal{A}_{j}\right\rangle =\frac{M-2}{2}%
\left\vert \mathcal{A}_{j}\right\rangle  \notag \\
\text{with \ \ }j &=&2,3,...,M+1  \label{4.10}
\end{eqnarray}

\item $l=2$ corresponds to the $\frac{\left( M-1\right) M}{2}$ states with
two particles with spin $\left\vert \Downarrow \right\rangle $. Again, we
will order these states in an arbitrary way:%
\begin{align}
\left\vert \mathcal{A}_{j}\right\rangle & =\left\vert \Uparrow ,\Uparrow
,,...,\Uparrow ,\Downarrow ,\Uparrow ,,...,\Uparrow ,\Downarrow ,\Uparrow
,...,\Uparrow ,\Uparrow \right\rangle \Longrightarrow H_{A}\left\vert 
\mathcal{A}_{j}\right\rangle =\frac{M-4}{2}\left\vert \mathcal{A}%
_{j}\right\rangle  \notag \\
\text{with \ \ }j& =M+2,M+3,...,M+1+\frac{\left( M-1\right) M}{2}
\label{4.11}
\end{align}

\item For the remaining values of $l$, the procedure is analogous.\medskip
\end{itemize}

Consequently, we have:%
\begin{align}
& 1\text{ eigenvector with eigenvalue }\frac{M}{2}  \notag \\
& M\text{ eigenvectors with eigenvalue }\frac{M-2}{2}  \notag \\
& \vdots  \notag \\
& \frac{M!}{(M-l)!l!}\text{ eigenvectors with eigenvalue }\frac{M-2l}{2}
\label{4.12}
\end{align}%
with $l=0,1,...M$. Then, it is clear that $H_{A}$ is degenerate: it has $%
2^{M}$ eigenvectors but only $M$ different eigenvalues. Therefore, a generic
state $\left\vert \mathcal{A}\right\rangle $ of the system $A$ can be
written in the basis $\left\{ \left\vert \mathcal{A}_{i}\right\rangle
\right\} $ as%
\begin{equation}
\left\vert \mathcal{A}\right\rangle =\sum_{i=1}^{2^{M}}C_{i}\left\vert 
\mathcal{A}_{i}\right\rangle \in \mathcal{H}_{A}\text{ \ \qquad\ with \ \ }%
\sum_{i=1}^{2^{M}}\left\vert C_{i}\right\vert ^{2}=1  \label{4.13}
\end{equation}%
By introducing eq. (\ref{4.13}) into eq. (\ref{4.6}), a pure initial state
of the composite system $U=A\cup B$ reads%
\begin{equation}
\left\vert \psi _{0}\right\rangle =\left( \sum_{i=1}^{2^{M}}C_{i}\left\vert 
\mathcal{A}_{i}\right\rangle \right) \otimes \left(
\bigotimes_{k=1}^{N}\left( \alpha _{k}\left\vert \uparrow _{k}\right\rangle
+\beta _{k}\left\vert \downarrow _{k}\right\rangle \right) \right)
\label{4.14}
\end{equation}%
If we group the degrees of freedom of $B$ in a single ket $\left\vert 
\mathcal{B}(0)\right\rangle $, $\left\vert \psi _{0}\right\rangle $ results%
\begin{equation}
|\psi _{0}\rangle =\sum_{i=1}^{2^{M}}C_{i}\left\vert \mathcal{A}%
_{i}\right\rangle \otimes \left\vert \mathcal{B}(0)\right\rangle
\label{4.15}
\end{equation}%
The time-evolution of $|\psi (t)\rangle $ is ruled by the time-evolution
operator $\mathcal{U}(t)=e^{-iHt}=e^{-i(H_{A}\otimes H_{B})t}$:%
\begin{equation}
|\psi (t)\rangle =\mathcal{U}(t)|\psi _{0}\rangle
=\sum_{i=1}^{2^{M}}C_{i}\,e^{-i(H_{A}\otimes H_{B})t}\,\left\vert \mathcal{A}%
_{i}\right\rangle \otimes \left\vert \mathcal{B}(0)\right\rangle
=\sum_{i=1}^{2^{M}}C_{i}\,e^{-iH_{A}t}\,\left\vert \mathcal{A}%
_{i}\right\rangle \otimes e^{-iH_{B}t}\,\left\vert \mathcal{B}%
(0)\right\rangle  \label{4.15-1}
\end{equation}%
If we use $\Lambda _{k}$ to denote the eigenvalue of $H_{A}$ corresponding
to the eigenvector $\left\vert \mathcal{A}_{k}\right\rangle $, then%
\begin{equation}
|\psi (t)\rangle =\sum_{i=1}^{2^{M}}C_{i}\,\left\vert \mathcal{A}%
_{i}\right\rangle \otimes e^{-i\Lambda _{i}H_{B}t}\,\left\vert \mathcal{B}%
(0)\right\rangle =\sum_{i=1}^{2^{M}}C_{i}\,\left\vert \mathcal{A}%
_{i}\right\rangle \otimes \left\vert \mathcal{B}(t)\right\rangle
\label{4.15-2}
\end{equation}%
where (see eq. (\ref{4.7}))%
\begin{equation}
\left\vert \mathcal{B}(t)\right\rangle =e^{-i\Lambda _{i}H_{B}t}\,\left\vert 
\mathcal{B}(0)\right\rangle =\exp \left[ -i\Lambda
_{k}\sum\limits_{j=1}^{N}g_{j}\left( \left\vert \uparrow _{j}\right\rangle
\left\langle \uparrow _{j}\right\vert -\left\vert \downarrow
_{j}\right\rangle \left\langle \downarrow _{j}\right\vert \right) t\right]
\left\vert \mathcal{B}(0)\right\rangle  \label{4.16}
\end{equation}%
Since the number of the eigenstates of $H_{A}$ with the same eigenvalue is
given by eqs. (\ref{4.12}), the terms of $|\psi (t)\rangle $ can be arranged
as%
\begin{align}
\left\vert \psi (t)\right\rangle & =\left( C_{1}\left\vert \mathcal{A}%
_{1}\right\rangle \left\vert \mathcal{B}_{0}(t)\right\rangle \right) +\left(
\sum\limits_{\lambda =1}^{M+1}C_{\lambda }\left\vert \mathcal{A}_{\lambda
}\right\rangle \left\vert \mathcal{B}_{1}(t)\right\rangle \right) +\left(
\sum\limits_{\lambda =M+2}^{M+1+\frac{\left( M-1\right) M}{2}}C_{\lambda
}\left\vert \mathcal{A}_{\lambda }\right\rangle \left\vert \mathcal{B}%
_{2}(t)\right\rangle \right) +...+  \notag \\
& +\left( \sum\limits_{\lambda =1+\sum_{p=0}^{l-1}\binom{M}{P}%
}^{\sum_{p=0}^{l}\binom{M}{P}}C_{\lambda }\left\vert \mathcal{A}_{\lambda
}\right\rangle \left\vert \mathcal{B}_{l}(t)\right\rangle \right)
+...+\left( C_{2^{M}}\left\vert \mathcal{A}_{2^{M}}\right\rangle \left\vert 
\mathcal{B}_{M}(t)\right\rangle \right)  \label{4.17}
\end{align}%
where%
\begin{equation}
\left\vert \mathcal{B}_{l}(t)\right\rangle
=\bigotimes\limits_{k=1}^{N}\left( \alpha _{k}e^{i\frac{\left( 2l-M\right) }{%
2}g_{k}t}\left\vert \uparrow _{k}\right\rangle +\beta _{k}e^{-i\frac{\left(
2l-M\right) }{2}g_{k}t}\left\vert \downarrow _{k}\right\rangle \right)
\label{4.18}
\end{equation}%
If we compare eq. (\ref{4.18}) with eq. (\ref{3.6}), we can see that $%
\left\vert \mathcal{E}_{\Uparrow }(t)\right\rangle $ and $\left\vert 
\mathcal{E}_{\Downarrow }(t)\right\rangle $ are the particular cases of $%
\left\vert \mathcal{B}_{l}(t)\right\rangle $ for $M=1$ and, then, $l=0,1$.
Let us recall that $l$ is the number of particles of the system $A$ having
spin $\left\vert \Downarrow \right\rangle $. Then, with $M=1$ and $l=0$, $%
\left\vert \mathcal{B}_{l}(t)\right\rangle =\left\vert \mathcal{E}_{\Uparrow
}(t)\right\rangle $, and with $M=1$ and $l=1$, $\left\vert \mathcal{B}%
_{l}(t)\right\rangle =\left\vert \mathcal{E}_{\Downarrow }(t)\right\rangle $.

If we define the function%
\begin{equation}
f(l)=\QATOPD\{ \} {\sum_{p=0}^{l}\binom{M}{P}\text{ if }l=0,1,...,M}{0\text{
\ \ \ \ \ \ \ \ \ \ otherwise}}  \label{4-19}
\end{equation}%
then eq. (\ref{4.17}) can be rewritten as%
\begin{equation}
\left\vert \psi (t)\right\rangle =\sum\limits_{l=0}^{M}\sum\limits_{\lambda
=f(l-1)+1}^{f(l)}C_{\lambda }\left\vert \mathcal{A}_{\lambda }\right\rangle
\left\vert \mathcal{B}_{l}(t)\right\rangle  \label{4.20}
\end{equation}%
and the state operator $\rho (t)=\left\vert \psi (t)\right\rangle
\left\langle \psi (t)\right\vert $ reads%
\begin{equation}
\rho (t)=\sum\limits_{l,l^{\prime }=0}^{M}\sum\limits_{\QATOP{\lambda
=f(l-1)+1}{\lambda ^{\prime }=f(l^{\prime }-1)+1}}^{\QATOP{f(l)}{f(l^{\prime
})}}C_{\lambda }C_{\lambda ^{\prime }}^{\ast }\left\vert \mathcal{A}%
_{\lambda }\right\rangle \left\vert \mathcal{B}_{l}(t)\right\rangle
\left\langle \mathcal{B}_{l^{\prime }}(t)\right\vert \left\langle \mathcal{A}%
_{\lambda ^{\prime }}\right\vert  \label{4.21}
\end{equation}

\subsection{Computing the expectation values}

An observable $O\in \mathcal{O}=\mathcal{H}\otimes \mathcal{H}$ of the
closed system $U=A\cup B$ can be expressed as%
\begin{equation}
O=\left( \sum\limits_{\lambda ,\lambda ^{\prime }=0}^{2^{M}}s_{\lambda
,\lambda ^{\prime }}\left\vert \mathcal{A}_{\lambda }\right\rangle
\left\langle \mathcal{A}_{\lambda ^{\prime }}\right\vert \right) \otimes
\left( \bigotimes_{i=1}^{N}\left( \epsilon _{\uparrow \uparrow
}^{(i)}\left\vert \uparrow _{i}\right\rangle \left\langle \uparrow
_{i}\right\vert +\epsilon _{\uparrow \downarrow }^{(i)}\left\vert \uparrow
_{i}\right\rangle \left\langle \downarrow _{i}\right\vert +\epsilon
_{\downarrow \uparrow }^{(i)}\left\vert \downarrow _{i}\right\rangle
\left\langle \uparrow _{i}\right\vert +\epsilon _{\downarrow \downarrow
}^{(i)}\left\vert \downarrow _{i}\right\rangle \left\langle \downarrow
_{i}\right\vert \right) \right)  \label{4.22}
\end{equation}%
Let us notice that eq. (\ref{3.7.1}) (a generic observable in the original
spin-bath model) is a particular case of this eq. (\ref{4.22}), with only
four terms in the first factor. Analogously to that case, the diagonal
components $s_{\lambda ,\lambda }$, $\epsilon _{\uparrow \uparrow }^{(i)}$, $%
\epsilon _{\downarrow \downarrow }^{(i)}$ are real numbers, and the
off-diagonal components are complex numbers satisfying $s_{\lambda ,\lambda
^{\prime }}=s_{\lambda ^{\prime },\lambda }^{\ast }$, $\epsilon _{\uparrow
\downarrow }^{(i)}=\epsilon _{\downarrow \uparrow }^{(i)\ast }$. Then, the
expectation value of the observable $O$ in the state $\rho (t)$ of eq. (\ref%
{4.21}) can be computed as%
\begin{equation}
\langle O\rangle _{\rho (t)}=Tr\left( O\rho (t)\right)
=\sum\limits_{l,l^{\prime }=0}^{M}\sum\limits_{\QATOP{\lambda =f(l-1)+1}{%
\lambda ^{\prime }=f(l^{\prime }-1)+1}}^{\QATOP{f(l)}{f(l^{\prime })}%
}B_{\lambda ,\lambda ^{\prime }}T_{l,l^{\prime }}(t)  \label{4.23}
\end{equation}%
where%
\begin{equation}
T_{l,l^{\prime }}(t)=\prod_{j=1}^{N}\left[ \left\vert \alpha _{j}\right\vert
^{2}\epsilon _{\uparrow \uparrow }^{(j)}e^{i\left( g_{j,l}-g_{j,l^{\prime
}}\right) \frac{t}{2}}+\left\vert \beta _{j}\right\vert ^{2}\epsilon
_{\downarrow \downarrow }^{(j)}e^{-i\left( g_{j,l}-g_{j,l^{\prime }}\right) 
\frac{t}{2}}+2\func{Re}\left( \alpha _{j}\beta _{j}^{\ast }\epsilon
_{\downarrow \uparrow }^{(j)}e^{i\left( g_{j,l}+g_{j,l^{\prime }}\right) 
\frac{t}{2}}\right) \right]  \label{4.24}
\end{equation}%
and 
\begin{equation}
g_{j,l}=\left( 2l-M\right) g_{j}\text{, \ \qquad\ }B_{\lambda ,\lambda
^{\prime }}=C_{\lambda }C_{\lambda ^{\prime }}^{\ast }s_{\lambda ^{\prime
},\lambda }  \label{4.25}
\end{equation}%
Since the exponents in eq. (\ref{4.24}) are of the form $g_{j,l}\pm
g_{j,l^{\prime }}$, in some cases they are zero. So, we can write%
\begin{equation}
\langle O\rangle _{\rho (t)}=\sum\limits_{l=0}^{M}\sum\limits_{\QATOP{%
\lambda =f(l-1)+1}{\lambda ^{\prime }=f(l-1)+1}}^{f(l)}B_{\lambda ,\lambda
^{\prime }}T_{l,l}(t)+\sum\limits_{l=0}^{\tilde{M}}\sum\limits_{\QATOP{%
\lambda =f(l-1)+1}{\lambda ^{\prime }=f(M-l-1)+1}}^{\QATOP{f(l)}{f(M-l)}%
}B_{\lambda ,\lambda ^{\prime }}2\func{Re}\left( T_{l,M-l}(t)\right)
+\sum\limits_{\QATOP{\QATOP{l,l^{\prime }=0}{l\neq l^{\prime }}}{l^{\prime
}\neq M-l}}^{M}\sum\limits_{\QATOP{\lambda =f(l-1)+1}{\lambda ^{\prime
}=f(l-1)+1}}^{\QATOP{f(l)}{f(l^{\prime })}}B_{\lambda ,\lambda ^{\prime
}}T_{l,l^{\prime }}(t)  \label{4.26}
\end{equation}%
where%
\begin{equation}
\tilde{M}=\QATOPD\{ \} {\frac{M-2}{2}\text{ if }M\text{ is even}}{\frac{M-1}{%
2}\text{\ if }M\text{ is odd}}  \label{4.27}
\end{equation}%
\begin{equation}
T_{l,l}(t)=\prod_{j=1}^{N}\left[ \left\vert \alpha _{j}\right\vert
^{2}\epsilon _{\uparrow \uparrow }^{(j)}+\left\vert \beta _{j}\right\vert
^{2}\epsilon _{\downarrow \downarrow }^{(j)}+2\func{Re}\left( \alpha
_{j}\beta _{j}^{\ast }\epsilon _{\downarrow \uparrow
}^{(j)}e^{ig_{j,l}t}\right) \right]  \label{4.28}
\end{equation}%
\begin{equation}
T_{l,M-l}(t)=\prod_{j=1}^{N}\left[ \left\vert \alpha _{j}\right\vert
^{2}\epsilon _{\uparrow \uparrow }^{(j)}e^{ig_{j,l}t}+\left\vert \beta
_{j}\right\vert ^{2}\epsilon _{\downarrow \downarrow }^{(j)}e^{-ig_{j,l}t}+2%
\func{Re}\left( \alpha _{j}\beta _{j}^{\ast }\epsilon _{\downarrow \uparrow
}^{(j)}\right) \right]  \label{4.29}
\end{equation}%
Let us notice that eqs. (\ref{4.28}) and (\ref{4.29}) are analogous to eqs. (%
\ref{3.9}) and (\ref{3.10}) for $\Gamma _{0}(t)$ and $\Gamma _{1}(t)$,
respectively, in the original model, with $g_{j,l}=\left( 2l-M\right) g_{j}$
instead of $g_{j}$. In particular, when $M=1$ and, so, $l=0,1$, then $%
T_{l,l}(t)=\Gamma _{0}(t)$ and $T_{l,M-l}(t)=\Gamma _{1}(t)$.

As in the case of the original spin-bath model, here we will consider
different meaningful ways of selecting the relevant observables.

\section{Generalized spin-bath model: Decomposition 1}

\subsection{Selecting the relevant observables}

In this case $A$ is the open system $S$ and $B$ is the environment $E$. This
is a generalization of Decomposition 1 in the original spin-bath model. The
only difference with respect to that case is that here the system $S$ is
composed of $M\geq 1$ particles instead of only one. Then, the TPS for this
case is%
\begin{equation}
\mathcal{H}=\mathcal{H}_{S}\otimes \mathcal{H}_{E}=\left(
\bigotimes\limits_{i=1}^{M}\mathcal{H}_{A_{i}}\right) \otimes \left(
\bigotimes\limits_{k=1}^{N}\mathcal{H}_{B_{k}}\right)  \label{5.0}
\end{equation}%
Therefore, the relevant observables $O_{R}$ of the closed system $U$ are
those corresponding to $A$, and they are obtained from eq. (\ref{4.22}) by
making $\epsilon _{\uparrow \uparrow }^{(i)}=\epsilon _{\downarrow
\downarrow }^{(i)}=1,$ $\epsilon _{\uparrow \downarrow }^{(i)}=0$ (compare
with eq. (\ref{3.11}) in the original spin-bath model):%
\begin{equation}
O_{R}=O_{S}\otimes \mathbb{I}_{E}=\left( \sum\limits_{\lambda ,\lambda
^{\prime }=0}^{2^{M}}s_{\lambda ,\lambda ^{\prime }}\left\vert \mathcal{A}%
_{\lambda }\right\rangle \left\langle \mathcal{A}_{\lambda ^{\prime
}}\right\vert \right) \otimes \left( \bigotimes_{i=1}^{N}\mathbb{I}%
_{i}\right)  \label{5.1}
\end{equation}%
With this condition, the expectation values of these observables are given
by eq. (\ref{4.26}), with%
\begin{eqnarray}
T_{l,l}(t) &=&\prod_{j=1}^{N}\left( \left\vert \alpha _{j}\right\vert
^{2}+\left\vert \beta _{j}\right\vert ^{2}\right) =1  \label{5.2} \\
T_{l,M-l}(t) &=&\prod_{j=1}^{N}\left( \left\vert \alpha _{j}\right\vert
^{2}e^{ig_{j,l}t}+\left\vert \beta _{j}\right\vert ^{2}e^{-ig_{j,l}t}\right)
\label{5.3} \\
T_{l,l^{\prime }}(t) &=&\prod_{j=1}^{N}\left( \left\vert \alpha
_{j}\right\vert ^{2}e^{i\left( g_{j,l}-g_{j,l^{\prime }}\right) \frac{t}{2}%
}+\left\vert \beta _{j}\right\vert ^{2}e^{-i\left( g_{j,l}-g_{j,l^{\prime
}}\right) \frac{t}{2}}\right)  \label{5.4}
\end{eqnarray}%
If we define the functions $R_{l}(t)=|T_{l,M-l}(t)|^{2}$ and $R_{ll^{\prime
}}(t)=|T_{l,l^{\prime }}(t)|^{2}$, they result%
\begin{eqnarray}
R_{l}(t) &=&\prod_{j=1}^{N}\left( \left\vert \alpha _{j}\right\vert
^{4}+\left\vert \beta _{j}\right\vert ^{4}+2\left\vert \alpha
_{j}\right\vert ^{2}\left\vert \beta _{j}\right\vert ^{2}\cos \left( 2\left(
2l-M\right) g_{j}t\right) \right)  \label{5.5} \\
R_{ll^{\prime }}(t) &=&\prod_{j=1}^{N}\left( \left\vert \alpha
_{j}\right\vert ^{4}+\left\vert \beta _{j}\right\vert ^{4}+2\left\vert
\alpha _{j}\right\vert ^{2}\left\vert \beta _{j}\right\vert ^{2}\cos \left(
2\left( l-l^{\prime }\right) g_{j}t\right) \right)  \label{5.6}
\end{eqnarray}%
We can see that $|r(t)|^{2}$ of eq. (\ref{3.14}) in the original model is
the particular case of $R_{l}(t)$ for $M=1$.

\subsection{Computing the behavior of the relevant expectation values}

The expectation value given by eq. (\ref{4.26}) has three terms, $\langle
O_{R}\rangle _{\rho (t)}=\Sigma ^{\left( 1\right) }+\Sigma ^{\left( 2\right)
}+\Sigma ^{\left( 3\right) }$, which can be analyzed separately:

\begin{itemize}
\item From eq. (\ref{5.2}), the first term reads%
\begin{equation}
\Sigma ^{\left( 1\right) }=\sum\limits_{l=0}^{M}\sum\limits_{\lambda
,\lambda ^{\prime }=f(l-1)+1}^{f(l)}B_{\lambda ,\lambda ^{\prime
}}=\sum\limits_{l=0}^{M}\sum\limits_{\lambda ,\lambda ^{\prime
}=f(l-1)+1}^{f(l)}C_{\lambda }C_{\lambda ^{\prime }}^{\ast }s_{\lambda
^{\prime },\lambda }\neq \Sigma ^{\left( 1\right) }(t)  \label{5.7}
\end{equation}%
It is clear that this first term does not evolve with time.

\item The time-dependence of the second term is given by $T_{l,M-l}(t)$:%
\begin{equation}
\Sigma ^{\left( 2\right) }(t)=\sum\limits_{l=0}^{\tilde{M}}\sum\limits_{%
\QATOP{\lambda =f(l-1)+1}{\lambda ^{\prime }=f(M-l-1)+1}}^{\QATOP{f(l)}{%
f(M-l)}}B_{\lambda ,\lambda ^{\prime }}2\func{Re}\left( T_{l,M-l}(t)\right)
\label{5.8}
\end{equation}%
Then, in order to obtain the limit of this term, we have to compute the
limit of $R_{l}(t)=|T_{l,M-l}(t)|^{2}$ of eq. (\ref{5.5}). As in the case of
the original spin-bath model, here we take $\left\vert \alpha
_{j}\right\vert ^{2}$ and $\left\vert \beta _{ji}\right\vert ^{2}$ as random
numbers in the closed interval $\left[ 0,1\right] $, such that $|\alpha
_{j}|^{2}+|\beta _{j}|^{2}=1$. Then%
\begin{eqnarray}
\max_{t}\left( \left\vert \alpha _{j}\right\vert ^{4}+\left\vert \beta
_{j}\right\vert ^{4}+2\left\vert \alpha _{j}\right\vert ^{2}\left\vert \beta
_{j}\right\vert ^{2}\cos \left( 2\left( 2l-M\right) g_{j}t\right) \right)
&=&1  \label{5.9} \\
\min_{t}\left( \left\vert \alpha _{j}\right\vert ^{4}+\left\vert \beta
_{j}\right\vert ^{4}+2\left\vert \alpha _{j}\right\vert ^{2}\left\vert \beta
_{j}\right\vert ^{2}\cos \left( 2\left( 2l-M\right) g_{j}t\right) \right)
&=&\left( 2\left\vert \alpha _{j}\right\vert ^{2}-1\right) ^{2}  \label{5.10}
\end{eqnarray}%
Therefore, $\left[ \left\vert \alpha _{j}\right\vert ^{4}+\left\vert \beta
_{j}\right\vert ^{4}+2\left\vert \alpha _{j}\right\vert ^{2}\left\vert \beta
_{j}\right\vert ^{2}\cos \left( 2\left( 2l-M\right) g_{j}t\right) \right] $
is a random number which, if $t\neq 0$, fluctuates between $1$ and $\left(
2\left\vert \alpha _{j}\right\vert ^{2}-1\right) ^{2}$. Again, when the
environment has many particles (that is, when $N\rightarrow \infty $), the
statistical value of the cases $\left\vert \alpha _{j}\right\vert ^{2}=1$, $%
\left\vert \beta _{j}\right\vert ^{2}=1$, $\left\vert \alpha _{j}\right\vert
^{2}=0$ and $\left\vert \beta _{j}\right\vert ^{2}=0$ tends to zero. In this
situation, eq. (\ref{5.5}) for $R_{l}(t)$ is an infinite product of numbers
belonging to the open interval $\left( 0,1\right) $. \ As a consequence,
when $N\rightarrow \infty $, $R_{l}(t)\rightarrow 0$.

\item The time-dependence of the third term is given by $T_{l,l^{\prime
}}(t) $:%
\begin{equation}
\Sigma ^{\left( 3\right) }(t)=\sum\limits_{\QATOP{\QATOP{l,l^{\prime }=0}{%
l\neq l^{\prime }}}{l^{\prime }\neq M-l}}^{M}\sum\limits_{\QATOP{\lambda
=f(l-1)+1}{\lambda ^{\prime }=f(l^{\prime }-1)+1}}^{\QATOP{f(l)}{f(l^{\prime
})}}B_{\lambda ,\lambda ^{\prime }}T_{l,l^{\prime }}(t)  \label{5.11}
\end{equation}%
with the restrictions on $l$ and $l^{\prime }$: $\ l\neq l^{\prime }$ and $%
l^{\prime }\neq M-l$. As in the second term, we have to compute the limit of 
$R_{ll^{\prime }}(t)=|T_{l,l^{\prime }}(t)|^{2}$ of eq. (\ref{5.6}) and, on
the basis of an analogous argument, the result is the same as above: when $%
N\rightarrow \infty $, $R_{ll^{\prime }}(t)\rightarrow 0$.
\end{itemize}

If we want now to evaluate the limit of $\langle O_{R}\rangle _{\rho (t)}$
for $t\rightarrow \infty $, we have to compute the limits of the second and
the third terms (since the first term, as we have seen, is
time-independent). Here we have to distinguish three cases: $M\ll N$, $M\gg
N $ and $M\simeq N$.\bigskip

\textbf{Case (a): }$M\ll N$

This case is similar to Decomposition 1 in the original spin-bath model,
since in both cases $M\ll N$: the only difference is that in the original
model $M=1$ whereas here $M\geq 1$.

In fact, we have seen that $T_{l,M-l}(t)$ is analogous to $\Gamma _{1}(t)$
in the original model. Moreover, $T_{l,l^{\prime }}(t)$ has the same
functional form as $\Gamma _{1}(t)$. In paper \cite{Max} it is shown that $%
\Gamma _{1}(t)$ approaches zero for $t\rightarrow \infty $. This means that
we can infer that $T_{l,M-l}(t)$ and $T_{l,l^{\prime }}(t)$ also approach
zero for $t\rightarrow \infty $. On the other hand, the terms $\Sigma
^{\left( 2\right) }(t)$ and $\Sigma ^{\left( 3\right) }(t)$ are sums of less
than $M$ terms involving $T_{l,M-l}(t)$ and $T_{l,l^{\prime }}(t)$. As a
consequence, since in this case $M$ is a small number, the sum of a small
number of terms approaching zero for $t\rightarrow \infty $ also approaches
zero: $\lim_{t\rightarrow \infty }\Sigma ^{\left( 2\right) }(t)=0$ and $%
\lim_{t\rightarrow \infty }\Sigma ^{\left( 3\right) }(t)=0$. Therefore,%
\begin{equation}
\lim_{t\rightarrow \infty }\langle O_{R}\rangle _{\rho
(t)}=\lim_{t\rightarrow \infty }\left[ \Sigma ^{\left( 1\right) }(t)+\Sigma
^{\left( 2\right) }(t)+\Sigma ^{\left( 3\right) }(t)\right] =\Sigma ^{\left(
1\right) }(t)  \label{5.11-1}
\end{equation}%
In other words, 
\begin{equation}
\lim_{t\rightarrow \infty }\langle O_{R}\rangle _{\rho
(t)}=\sum\limits_{l=0}^{M}\sum\limits_{\lambda ,\lambda ^{\prime
}=f(l-1)+1}^{f(l)}B_{\lambda ,\lambda ^{\prime
}}=\sum\limits_{l=0}^{M}\sum\limits_{\lambda ,\lambda ^{\prime
}=f(l-1)+1}^{f(l)}C_{\lambda }C_{\lambda ^{\prime }}^{\ast }s_{\lambda
^{\prime },\lambda }=\langle O_{R}\rangle _{\rho _{\ast }}  \label{5.12}
\end{equation}%
where $\rho _{\ast }$ is the final diagonal state of $U$. This result can
also be expressed in terms of the reduced density operator $\rho _{A}$ of
the system $A$ as (see eq. (\ref{2.5})):%
\begin{equation}
\lim_{t\rightarrow \infty }\langle O_{R}\rangle _{\rho (t)}=\langle
O_{R}\rangle _{\rho _{\ast }}=\lim_{t\rightarrow \infty }\langle
O_{A}\rangle _{\rho _{A}(t)}=\langle O_{A}\rangle _{\rho _{A\ast }}
\label{5.13}
\end{equation}%
In the eigenbasis of the Hamiltonian $H_{A}$ of $A$, the final reduced
density operator $\rho _{A\ast }$ is expressed by a $2^{M}\times 2^{M}$\
matrix:%
\begin{equation}
\rho _{A\ast }=%
\begin{pmatrix}
\rho _{l=0} & 0 & 0 & 0 & ... & 0 \\ 
0 & \rho _{l=1} & 0 & 0 & ... & 0 \\ 
0 & 0 & \rho _{l=2} & 0 & ... & 0 \\ 
0 & 0 & 0 & \rho _{l=3} & ... & 0 \\ 
... & ... & ... & ... & ... & ... \\ 
0 & 0 & 0 & 0 & 0 & \rho _{l=M}%
\end{pmatrix}
\label{5.14}
\end{equation}%
where $\rho _{l=0}=\left\vert C_{1}\right\vert ^{2}$ and each $\rho _{l}$ is
a matrix of dimension $\frac{M!}{\left( M-l\right) !l!}\times \frac{M!}{%
\left( M-l\right) !l!}$.\ This result might seem insufficient for
decoherence because, since the $\rho _{l}$ are matrices, $\rho _{A\ast }$
seems to be non completely diagonal in the eigenbasis of the Hamiltonian $%
H_{A}$. However, we have to recall that all the states $\left\vert \mathcal{A%
}_{i}\right\rangle $ with same $l$ are degenerate eigenvectors corresponding
to the same eigenvalue of $H_{A}$; then, the basis that diagonalizes $\rho
_{A\ast }$ (i.e., that diagonalizes all the matrices $\rho _{l}$) is an
eigenbasis of $H_{A}$. Summing up, the system $S=A$ of $M$ particles in
interaction with its environment $E=B$ of $N\gg M$ particles decoheres in
the eigenbasis of $\rho _{A\ast }$, which is also an eigenbasis of $H_{A}$.

If we want to compute the time-behavior of $\langle O_{R}\rangle _{\rho (t)}$%
, we have to consider that $\Sigma ^{\left( 1\right) }$ is a sum of terms of
the form $(B_{\lambda ,\lambda ^{\prime }}\left\vert \alpha _{j}\right\vert
^{2}+B_{\lambda ,\lambda ^{\prime }}\left\vert \beta _{j}\right\vert ^{2})$,
that is, terms of the expectation value coming from the diagonal part of $%
\rho (t)$ in the basis of the Hamiltonian $H$. Therefore, if there is
decoherence, the sum $\Sigma ^{nd}(t)=\Sigma ^{\left( 2\right) }+\Sigma
^{\left( 3\right) }$, involving the terms of $\langle O_{R}\rangle _{\rho
(t)}$ coming from the non-diagonal part of $\rho (t)$, has to approach zero
for $t\rightarrow \infty $.

In order to show an example of the time-behavior of $\langle O_{R}\rangle
_{\rho (t)}$, numerical simulations for $\Sigma ^{nd}(t)$ have been
performed, with the following features:

\begin{enumerate}
\item[(i)] $s_{\lambda ^{\prime },\lambda }=1$ (see eq. (\ref{5.1})).

\item[(ii)] The initial condition for $S=A$ is selected as (see eq. (\ref%
{4.13})) : 
\begin{equation}
\left\vert \mathcal{A}\right\rangle =\frac{1}{\sqrt{2^{M}}}%
\sum_{i=1}^{2^{M}}\left\vert \mathcal{A}_{i}\right\rangle \Longrightarrow
\forall \lambda ,\ C_{\lambda }=C_{\lambda }^{\ast }=\frac{1}{\sqrt{2^{M}}}%
\Longrightarrow C_{\lambda }C_{\lambda ^{\prime }}^{\ast }=\frac{1}{2^{M}}
\end{equation}%
Then, from (i) and (ii), $B_{\lambda ,\lambda ^{\prime }}=2^{-M}$ (see eq. (%
\ref{4.25})).

\item[(iii)] $\left\vert \alpha _{i}\right\vert ^{2}$ is generated by a
random-number generator in the interval $[0,1]$, and $\left\vert \beta
_{i}\right\vert ^{2}$ is obtained as $\left\vert \beta _{i}\right\vert
^{2}=1-\left\vert \alpha _{i}\right\vert ^{2}$.

\item[(iv)] $g_{i}=400Hz$: as explained above, the coupling constant in
typical models of spin interaction.

\item[(v)] As in the original model, the time-interval $\left[ 0,t_{0}\right]
$ was partitioned into intervals $\Delta t=t_{0}/200$, and the function $%
\Sigma ^{nd}(t)$ was computed at times $t_{k}=k\Delta t$, with $%
k=0,1,...,200 $.

\item[(vi)] $N=10^{3}$, and $M=1$ and $M=10$.
\end{enumerate}

Figure 7 shows the time-evolution of $\Sigma ^{nd}(t)$.%

 \begin{figure}[t]

 \centerline{\scalebox{0.7}{\includegraphics{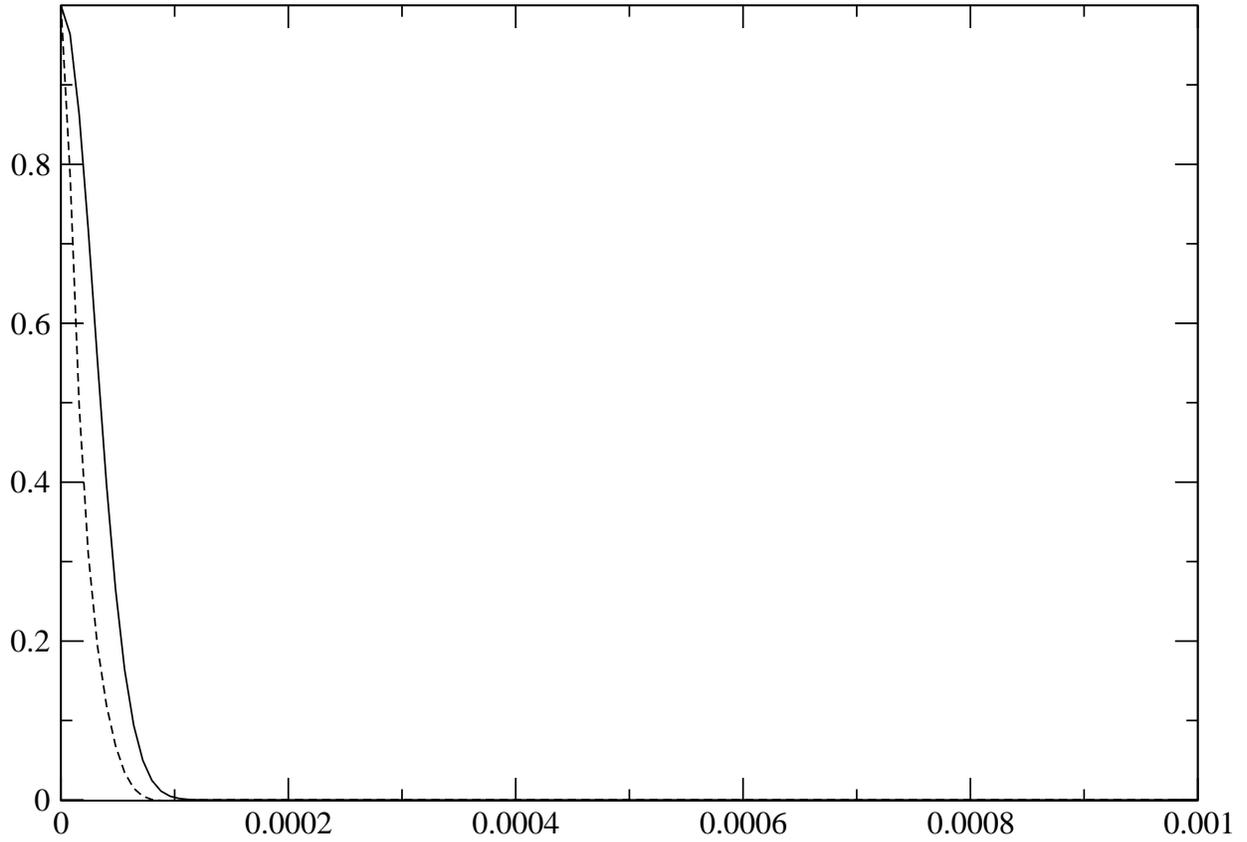}}}
\caption{Figure 7: Evolution of $%
\Sigma ^{nd}(t)$ for $N=10^{3}$, and $M=1$ (solid line) and $M=10$ (dash
line), with $t_{0}=10^{-3}s$.}
 \label{f1}\vspace*{0.cm}
\end{figure}

This result shows that, as expected, a small open system $S=A$ of $M$
particles decoheres in interaction with a large environment $E=B$ of $N\gg M$
particles.\bigskip 

\textbf{Case (b): }$M\gg N$

In this case, where the open system $S=$ $A$ has much more particles than
the environment $E=B$, the argument of Case (a) cannot be applied: since now 
$\Sigma ^{\left( 2\right) }(t)$ and $\Sigma ^{\left( 3\right) }(t)$ are no
longer sums over a small number of terms, the fact that each term approaches
zero does not guarantee that the sums also approach zero. In particular, if $%
N=1$, then (see eq. (\ref{5.4}))%
\begin{equation}
T_{l,l^{\prime }}(t)=\left\vert \alpha _{1}\right\vert ^{2}e^{i\left(
g_{1,l}-g_{1,l^{\prime }}\right) \frac{t}{2}}+\left\vert \beta
_{1}\right\vert ^{2}e^{-i\left( g_{1,l}-g_{1,l^{\prime }}\right) \frac{t}{2}}
\label{5.16}
\end{equation}%
which clearly has no limit for $t\rightarrow \infty $. Nevertheless, it
might happen that, with $N$ high but $M$ much higher than $N$, each term of
the sums approaches zero. So, in order to know the time behavior of $\langle
O_{R}\rangle _{\rho (t)}$, numerical simulations for $\Sigma ^{nd}(t)$ have
been performed, with the same features as in the previous case, with the
exception of condition (vi), which was taken as:

\begin{enumerate}
\item[(vi)] $M=10^{3}$, and $N=10$ and $N=100$.
\end{enumerate}

Figure 8 shows the time-evolution of $\Sigma ^{nd}(t)$ in this case.%

 \begin{figure}[t]

 \centerline{\scalebox{0.7}{\includegraphics{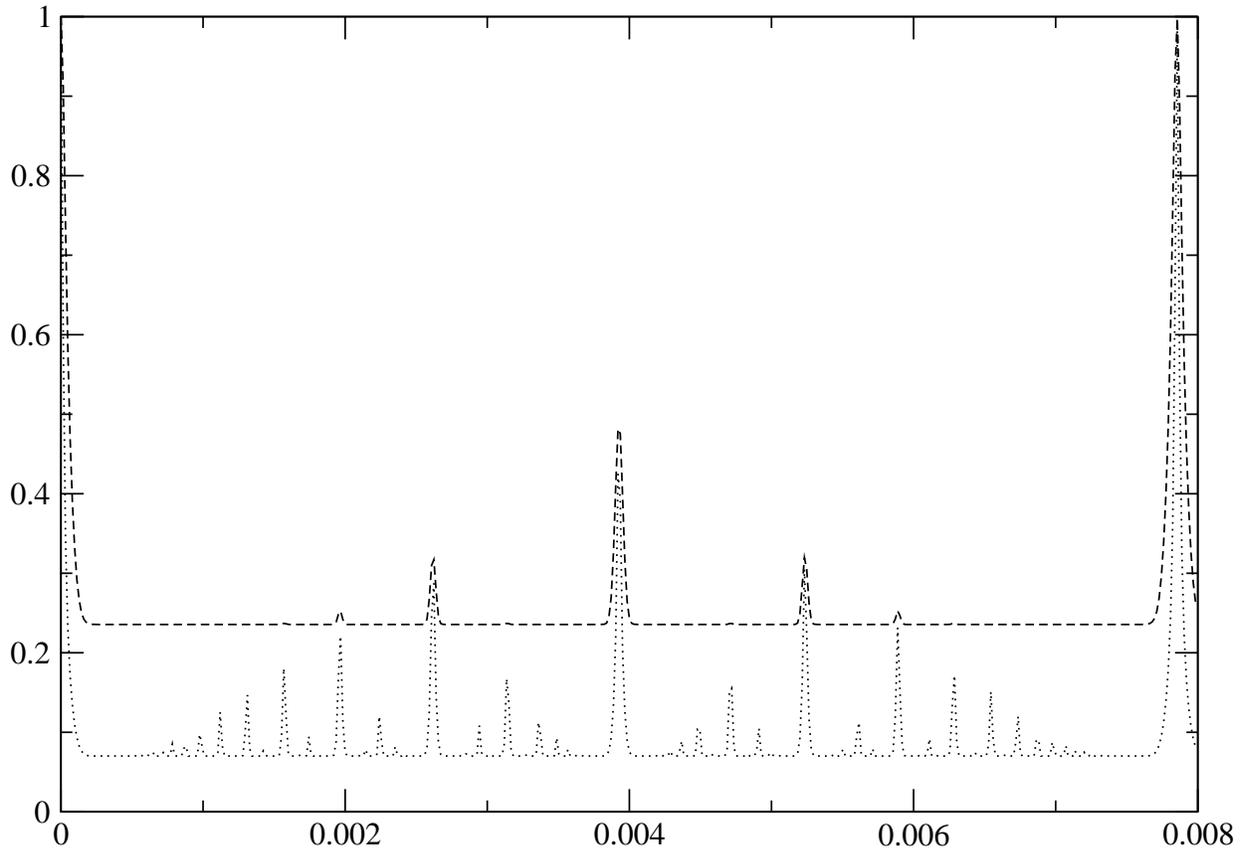}}}
\caption{Figure 8: Evolution of $%
\Sigma ^{nd}(t)$ for $M=10^{3}$, and $N=10$ (dash line) and $N=100$ (dot
line), with $t_{0}=10^{-3}s$.}
 \label{f1}\vspace*{0.cm}
\end{figure}

This result is also what may be expected from the EID perspective: when \
the open system $S=A$ of $M$ particles is larger that the environment $E=B$
of $N\ll M$ particles, $S$ does not decohere.\bigskip 

\textbf{Case (c): }$M\simeq N$

In this case, where the numbers of particles of the open system $S=$ $A$ and
of the environment $E=B$ do not differ in more than one order of magnitude,
the time behavior of $\langle O_{R}\rangle _{\rho (t)}$ cannot be inferred
from the equations. Numerical simulations have been performed, with the same
features as in Case (b), with the exception of condition (vi), which was
taken as:

\begin{enumerate}
\item[(vi)] $N=10^{3}$, and $M=10^{2}$ and $M=10^{3}$.
\end{enumerate}

Figure 9 shows the time-evolution of $\Sigma ^{nd}(t)$.%

 \begin{figure}[t]

 \centerline{\scalebox{0.7}{\includegraphics{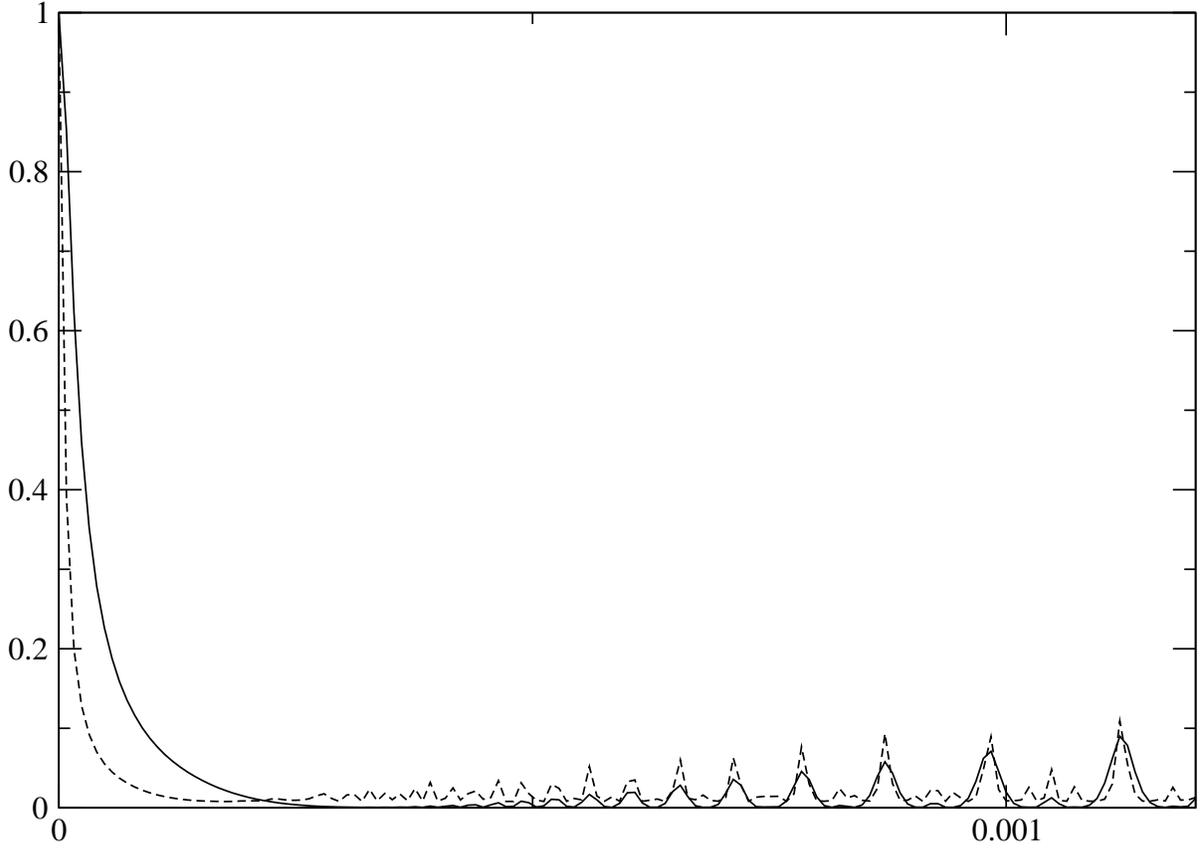}}}
\caption{Figure 9: Evolution of $%
\Sigma ^{nd}(t)$ for $N=10^{3}$, and $M=10^{2}$ (dash line) and $M=10^{3}$
(solid line), with $t_{0}=12.10^{-4}s$.}
 \label{f1}\vspace*{0.cm}
\end{figure}

Again, this result is not surprising from the viewpoint of the EID approach:
if the environment $E=B$ of $N$ particles is not large enough when compared
with the open system $S=A$ of $M$ particles, $S$ does not decohere.

Let us notice that, for $N=10^{3}$, the system $S=A$ with $M=10^{2}$ does
not decohere (Figure 9), whereas it does decohere with $M=10$ (Figure 7).
This shows that, in the case of this decomposition, $M\ll N$ means that $N$
is at least two orders of magnitude higher than $M$.\bigskip

\textbf{Summarizing results}

Up to now, in this Decomposition 1 all the arguments were directed to know
whether the system $A$ of $M$ particles decoheres or not in interaction with
the system $B$ of $N$ particles. But, given the symmetry of the whole
system, the same arguments can be used to decide whether the system $B$ of $%
N $ particles decoheres or not in interaction with the system $A$ of $M$
particles, with analogous results: $B$ decoheres only when $M\gg N$; if $%
M\ll N$ or $M\simeq N$, $B$ does not decohere. Therefore, all the results
obtained in this section can be summarized as follows:

\begin{enumerate}
\item[(i)] If $M\ll N$, $A$ decoheres and $B$ does not decohere.

\item[(ii)] If $M\gg N$, $A$ does not decohere and $B$ decoheres.

\item[(iii)] If $M\simeq N$, neither $A$ nor $B$ decohere.
\end{enumerate}

This general conclusion completely agrees with the usual reading of the EID
approach: the decoherence of an open system $S$ is the result of its
interaction with a very large environment $E$. This happens when $S=A$ in
(i) and when $S=B$ in (ii), but it does not happen in (iii).

\section{Generalized spin-bath model: Decomposition 2}

\subsection{Selecting the relevant observables}

In this case we decide to observe only one particle of the open system $A$.
This amounts to splitting the closed system $U$ into two new subsystems: the
open system $S$ is, say, the particle $A_{M}$ with ket $\left\vert \Uparrow
,\Uparrow ,...,\Uparrow ,\Uparrow ,\Uparrow ,\Downarrow \right\rangle $, and
the environment is $E=\left( \cup _{i=1}^{M-1}A_{i}\right) \cup B=\left(
\cup _{i=1}^{M-1}A_{i}\right) \cup \left( \cup _{k=1}^{N}B_{k}\right) $. Let
us notice that the Decomposition 2 of the original spin-bath model is a
particular case of this one, for $N=1$ (see eq. (\ref{4-0}), where $N$ plays
the role of the $M$ of this case). The TPS for this case is%
\begin{equation}
\mathcal{H}=\mathcal{H}_{S}\otimes \mathcal{H}_{E}=\left( \mathcal{H}%
_{A_{M}}\right) \otimes \left( \left( \bigotimes\limits_{i=1}^{M-1}\mathcal{H%
}_{A_{i}}\right) \otimes \left( \bigotimes\limits_{k=1}^{N}\mathcal{H}%
_{B_{k}}\right) \right)  \label{6.0}
\end{equation}%
Therefore, the relevant observables $O_{R}$ of the closed system $U$ are
those corresponding to the particle $A_{M}$:%
\begin{equation}
O_{R}=O_{S}\otimes \mathbb{I}_{E}=\left( \sum\limits_{\alpha ,\alpha
^{\prime }=\Uparrow ,\Downarrow }s_{\alpha ,\alpha ^{\prime }}\left\vert
\alpha \right\rangle \left\langle \alpha ^{\prime }\right\vert \right)
\otimes \left( \left( \bigotimes_{i=1}^{M-1}\mathbb{I}_{i}\right) \otimes
\left( \bigotimes_{k=1}^{N}\mathbb{I}_{k}\right) \right)  \label{6.1}
\end{equation}%
It is easy to see that the relevant observables selected in this
Decomposition 2 form a subspace of the space of the relevant observables
selected in Decomposition 1: eq. (\ref{6.1}) can be obtained from eq. (\ref%
{5.1}) by making $s_{\lambda ,\lambda ^{\prime }}=1$ for $\lambda =\lambda
^{\prime }$ and $s_{\lambda ,\lambda ^{\prime }}=0$ for $\lambda \neq
\lambda ^{\prime }$ in all the terms of the sum except for the terms
corresponding to the particle $A_{M}$.

In order to simplify expressions, in this case it is convenient to introduce
a new arrangement for the eigenvectors of the Hamiltonian $H_{A}$, by
calling them $\left\vert \mathcal{\tilde{A}}_{i}\right\rangle $: the set $%
\left\{ \left\vert \mathcal{\tilde{A}}_{i}\right\rangle \right\} $ is an
eigenbasis of $H_{A}$ with $2^{M}$ elements. The $\left\vert \mathcal{\tilde{%
A}}_{i}\right\rangle $ will be ordered by analogy with the binary numbers:%
\begin{eqnarray}
\left\vert \mathcal{\tilde{A}}_{1}\right\rangle &=&\left\vert \Uparrow
,\Uparrow ,...,\Uparrow ,\Uparrow ,\Uparrow ,\Uparrow \right\rangle \text{, }%
\left\vert \mathcal{\tilde{A}}_{2}\right\rangle =\left\vert \Uparrow
,\Uparrow ,...,\Uparrow ,\Uparrow ,\Uparrow ,\Downarrow \right\rangle \text{%
, }\left\vert \mathcal{\tilde{A}}_{3}\right\rangle =\left\vert \Uparrow
,\Uparrow ,...,\Uparrow ,\Uparrow ,\Downarrow ,\Uparrow \right\rangle \text{,%
}  \notag \\
\text{ }\left\vert \mathcal{\tilde{A}}_{4}\right\rangle &=&\left\vert
\Uparrow ,\Uparrow ,...,\Uparrow ,\Uparrow ,\Downarrow ,\Downarrow
\right\rangle \text{, }\left\vert \mathcal{\tilde{A}}_{5}\right\rangle
=\left\vert \Uparrow ,\Uparrow ,...,\Uparrow ,\Downarrow ,\Uparrow ,\Uparrow
\right\rangle \text{, }\left\vert \mathcal{\tilde{A}}_{6}\right\rangle
=\left\vert \Uparrow ,\Uparrow ,...,\Uparrow ,\Downarrow ,\Uparrow
,\Downarrow \right\rangle ,...  \notag \\
\left\vert \mathcal{\tilde{A}}_{2^{M}}\right\rangle &=&\left\vert \Downarrow
,\Downarrow ,...,\Downarrow ,\Downarrow ,\Downarrow ,\Downarrow \right\rangle
\label{6.2}
\end{eqnarray}%
According to this arrangement, the $\left\vert \mathcal{\tilde{A}}%
_{i}\right\rangle $ with even $i$ have the spin $M$ in the state $\left\vert
\Downarrow \right\rangle $, and the $\left\vert \mathcal{\tilde{A}}%
_{i}\right\rangle $ with odd $i$ have the spin $M$ in the state $\left\vert
\Uparrow \right\rangle $. So, the relevant observables of eq. (\ref{6.1})
can be rewritten in terms of the $\left\vert \mathcal{\tilde{A}}%
_{i}\right\rangle $ as%
\begin{equation}
O_{R}=\left( \sum\limits_{\lambda =1}^{2^{M}}\left( \tilde{s}_{\Uparrow
\Uparrow }\left\vert \mathcal{\tilde{A}}_{2\lambda }\right\rangle
\left\langle \mathcal{\tilde{A}}_{2\lambda }\right\vert +\tilde{s}_{\Uparrow
\Downarrow }\left\vert \mathcal{\tilde{A}}_{2\lambda }\right\rangle
\left\langle \mathcal{\tilde{A}}_{2\lambda -1}\right\vert +\tilde{s}%
_{\Downarrow \Uparrow }\left\vert \mathcal{\tilde{A}}_{2\lambda
-1}\right\rangle \left\langle \mathcal{\tilde{A}}_{2\lambda }\right\vert +%
\tilde{s}_{\Downarrow \Downarrow }\left\vert \mathcal{\tilde{A}}_{2\lambda
-1}\right\rangle \left\langle \mathcal{\tilde{A}}_{2\lambda -1}\right\vert
\right) \right) \otimes \left( \bigotimes_{k=1}^{N}\mathbb{I}_{k}\right)
\label{6.3}
\end{equation}

\subsection{Computing the behavior of the relevant expectation values}

Here the expectation values of the relevant observables are given by eq. (%
\ref{4.26}), with $T_{l,l^{\prime }}(t)$, $T_{l,l}(t)$ and $T_{l,M-l}(t)$
given by eqs. (\ref{4.24}), (\ref{4.28}) and (\ref{4.29}) respectively, but
now replacing $B_{\lambda ,\lambda ^{\prime }}$ with $\tilde{B}_{\lambda
,\lambda ^{\prime }}$, 
\begin{equation}
\langle O_{R}\rangle _{\rho (t)}=\sum\limits_{l=0}^{M}\sum\limits_{\QATOP{%
\lambda =f(l-1)+1}{\lambda ^{\prime }=f(l-1)+1}}^{f(l)}\tilde{B}_{\lambda
,\lambda ^{\prime }}+\sum\limits_{l=0}^{\tilde{M}}\sum\limits_{\QATOP{%
\lambda =f(l-1)+1}{\lambda ^{\prime }=f(M-l-1)+1}}^{\QATOP{f(l)}{f(M-l)}}%
\tilde{B}_{\lambda ,\lambda ^{\prime }}2\func{Re}\left( T_{l,M-l}(t)\right)
+\sum\limits_{\QATOP{\QATOP{l,l^{\prime }=0}{l\neq l^{\prime }}}{l^{\prime
}\neq M-l}}^{M}\sum\limits_{\QATOP{\lambda =f(l-1)+1}{\lambda ^{\prime
}=f(l^{\prime }-1)+1}}^{\QATOP{f(l)}{f(l^{\prime })}}\tilde{B}_{\lambda
,\lambda ^{\prime }}T_{l,l^{\prime }}(t)  \label{6.4}
\end{equation}%
where the $\tilde{B}_{\lambda ,\lambda ^{\prime }}$ can be written in the
basis $\left\{ \left\vert \mathcal{\tilde{A}}_{\lambda }\right\rangle
\right\} $ as%
\begin{equation}
\tilde{B}_{\lambda ,\lambda ^{\prime }}=\left\{ 
\begin{array}{ccccc}
C_{\lambda }C_{\lambda ^{\prime }}^{\ast }\tilde{s}_{\Uparrow \Uparrow } & 
\text{if} & \lambda & \text{is an even number and} & \lambda ^{\prime
}=\lambda \\ 
C_{\lambda }C_{\lambda ^{\prime }}^{\ast }\tilde{s}_{\Uparrow \Downarrow } & 
\text{if} & \lambda & \text{is an even number and} & \lambda ^{\prime
}=\lambda -1 \\ 
C_{\lambda }C_{\lambda ^{\prime }}^{\ast }\tilde{s}_{\Downarrow \Uparrow } & 
\text{if} & \lambda & \text{is an odd number and} & \lambda ^{\prime
}=\lambda +1 \\ 
C_{\lambda }C_{\lambda ^{\prime }}^{\ast }\tilde{s}_{\Downarrow \Downarrow }
& \text{if} & \lambda & \text{is an odd number and} & \lambda ^{\prime
}=\lambda \\ 
0 &  &  & \text{otherwise} & 
\end{array}%
\right\}  \label{6.5}
\end{equation}%
According to eq. (\ref{6.5}), $\tilde{B}_{\lambda ,\lambda ^{\prime }}\neq 0$
only when

\begin{equation}
\lambda ^{\prime }=\lambda \text{ \ \ \ \ or \ \ \ }\lambda ^{\prime
}=\lambda \pm 1\text{ }  \label{6.5b}
\end{equation}%
Since $\lambda =f(l-1)+1$ and $\lambda ^{\prime }=f(l^{\prime }-1)+1$,
relations (\ref{6.5b}) imply that 
\begin{equation}
l^{\prime }=l\text{ \ \ \ \ or \ \ \ }l^{\prime }=l\pm 1  \label{6.5c}
\end{equation}

The expectation value given by eq. (\ref{6.4}) has again three terms, $%
\langle O\rangle _{\rho (t)}=\Sigma ^{\left( 1\right) }+\Sigma ^{\left(
2\right) }+\Sigma ^{\left( 3\right) }$, which can be analyzed separately:

\begin{itemize}
\item From eqs. (\ref{6.5}) and (\ref{6.5b}), the first term reads%
\begin{equation}
\Sigma ^{\left( 1\right) }=\sum\limits_{l=0}^{M}\sum\limits_{\lambda
=f(l-1)+1}^{f(l)}B_{\lambda ,\lambda }=\sum\limits_{\lambda
=0}^{2^{M-1}}\left( \left\vert C_{2\lambda }\right\vert ^{2}\tilde{s}%
_{\Uparrow \Uparrow }+\left\vert C_{2\lambda +1}\right\vert ^{2}\tilde{s}%
_{\Downarrow \Downarrow }\right) \neq \Sigma ^{\left( 1\right) }(t)
\label{6.5d}
\end{equation}%
Analogously to eq. (\ref{5.7}) of Decomposition 1, this first term does not
evolve with time.

\item The time-dependence of the second term is given by $T_{l,M-l}(t)$. But
with the restrictions of eqs. (\ref{6.5b}) and (\ref{6.5c}), $\Sigma
^{\left( 2\right) }$ has only two terms:%
\begin{eqnarray}
\Sigma ^{\left( 2\right) }(t) &=&\sum\limits_{l=0}^{\tilde{M}}\sum\limits_{%
\QATOP{\lambda =f(l-1)+1}{\lambda ^{\prime }=f(M-l-1)+1}}^{\QATOP{f(l)}{%
f(M-l)}}B_{\lambda ,\lambda ^{\prime }}2\func{Re}\left( T_{l,M-l}(t)\right) =
\\
&=&C_{f(\frac{M-1}{2}-1)+1}C_{f(\frac{M-1}{2}-1)+2}^{\ast }\left( \tilde{s}%
_{\Downarrow \Uparrow }+\tilde{s}_{\Uparrow \Downarrow }\right) 2\func{Re}%
\left( T_{\frac{M-1}{2},\frac{M+1}{2}}(t)\right)  \label{6.5.e}
\end{eqnarray}%
Then, in order to obtain the limit of this term, we have to compute the
limit of $T_{\frac{M-1}{2},\frac{M+1}{2}}(t)$, which is precisely the $%
T_{l,l^{\prime }}(t)$ of Decomposition 1 in the particular case that $l=%
\frac{M-1}{2}$ and $l^{\prime }=\frac{M+1}{2}$ (see eq. (\ref{5.4})). But,
as we have seen in Case (a) of Decomposition 1, $T_{l,l^{\prime }}(t)$ has
the same functional form as $\Gamma _{1}(t)$ of the original model (see eq. (%
\ref{3.10})), which approaches zero for $t\rightarrow \infty $ when $N\gg 1$%
. Therefore, for $N\gg 1$, $T_{\frac{M-1}{2},\frac{M+1}{2}}(t)$ also
approaches zero for $t\rightarrow \infty $, and the same holds for $\Sigma
^{\left( 2\right) }(t)$ since it is a sum of two terms containing $T_{\frac{%
M-1}{2},\frac{M+1}{2}}(t)$.

\item The time-dependence of the third term is given by $T_{l,l^{\prime
}}(t) $. But with the restrictions of eqs. (\ref{6.5b}) and (\ref{6.5c}), $%
\Sigma ^{\left( 3\right) }$ results:%
\begin{equation}
\Sigma ^{\left( 3\right) }(t)=\sum\limits_{\QATOP{l=0}{l\neq \frac{M-1}{2}}%
}^{M}\sum\limits_{\lambda =f(l-1)+1}^{f(l)}\left( B_{\lambda ,\lambda
+1}T_{l,l+1}(t)+B_{\lambda ,\lambda -1}T_{l,l-1}(t)\right)  \label{6.5h}
\end{equation}%
Since here $l^{\prime }=l\pm 1$ (see eq. (\ref{6.5c})), in this case $%
T_{l,l\pm 1}(t)$ is:%
\begin{equation}
T_{l,l\pm 1}(t)=\prod_{j=1}^{N}\left( \left\vert \alpha _{j}\right\vert
^{2}e^{\mp ig_{j}t}+\left\vert \beta _{j}\right\vert ^{2}e^{\pm
ig_{j}t}\right)  \label{6.5i}
\end{equation}%
If we compare this equation with eq. (\ref{3.13}) for $r(t)$ in the original
spin-bath model, we can see that 
\begin{equation}
T_{l,l+1}(t)=r(t)\text{ \ \ \ \ and \ \ \ }T_{l,l-1}(t)=r^{\ast }(t)
\label{6.5j}
\end{equation}%
Then, 
\begin{equation}
\Sigma ^{\left( 3\right) }(t)=\left( S_{+}r(t)+S_{-}r^{\ast }(t)\right)
\label{6.5k}
\end{equation}%
where $S_{+}$ and $S_{-}$ are constants given by 
\begin{equation}
S_{\pm }=\sum\limits_{\QATOP{l=0}{l\neq \frac{M-1}{2}}}^{M}\sum\limits_{%
\lambda =f(l-1)+1}^{f(l)}B_{\lambda ,\lambda \pm 1}  \label{6.5m}
\end{equation}%
On the basis of the simulations of the original model we have seen that,
when $N\gg 1$, $r(t)$ approaches zero for $t\rightarrow \infty $. Therefore,
in this case we can conclude that, when $N\gg 1$, $\Sigma ^{\left( 3\right)
}(t)$ approaches zero for $t\rightarrow \infty $.
\end{itemize}

Summing up, $\langle O_{R}\rangle _{\rho (t)}$ is the sum of three terms:
one is time-independent and the other two tend to zero for $t\rightarrow
\infty $. In particular, from eq. (\ref{6.5d}) we know that, for $N\gg 1$,

\begin{equation}
\lim_{t\rightarrow \infty }\langle O_{R}\rangle _{\rho
(t)}=\sum\limits_{l=0}^{M}\sum\limits_{\lambda ,\lambda ^{\prime
}=f(l-1)+1}^{f(l)}\tilde{B}_{\lambda ,\lambda ^{\prime
}}=\sum\limits_{l=0}^{M}\sum\limits_{\lambda =0}^{2^{M-1}}\left( \left\vert
C_{2\lambda }\right\vert ^{2}\tilde{s}_{\Uparrow \Uparrow }+\left\vert
C_{2\lambda +1}\right\vert ^{2}\tilde{s}_{\Downarrow \Downarrow }\right)
=\langle O_{R}\rangle _{\rho _{\ast }}  \label{6.6}
\end{equation}%
where $\rho _{\ast }$ is the final diagonal state of $U$. Again, this result
can also be expressed in terms of the reduced density operator $\rho
_{S}=\rho _{A_{M}}$ of the open system $S=A_{M}$ as (see eq. (\ref{5.13}))%
\begin{equation}
\lim_{t\rightarrow \infty }\langle O_{R}\rangle _{\rho (t)}=\langle
O_{R}\rangle _{\rho _{\ast }}=\lim_{t\rightarrow \infty }\langle
O_{A_{M}}\rangle _{\rho _{A_{M}}(t)}=\langle O_{A_{M}}\rangle _{\rho
_{A_{M}\ast }}  \label{6.7}
\end{equation}%
where the final reduced density operator $\rho _{A_{M}\ast }$ in the basis $%
\{\left\vert \Uparrow \right\rangle ,\left\vert \Downarrow \right\rangle \}$
reads%
\begin{equation}
\rho _{A_{M}\ast }=%
\begin{pmatrix}
\left\vert \alpha _{M}\right\vert ^{2} & 0 \\ 
0 & \left\vert \beta _{M}\right\vert ^{2}%
\end{pmatrix}
\label{6.8}
\end{equation}%
This shows that the open system $S=A_{M}$, composed of a single particle,
decoheres in interaction with its environment $E$ of $N+M-1$ particles when $%
N\gg 1$, independently of the value of $M$.\bigskip

In order to illustrate this conclusion, we have computed $\Sigma
^{nd}(t)=\Sigma ^{\left( 2\right) }(t)+\Sigma ^{\left( 3\right) }(t)$ by
means of numerical simulations with the same features as in Decomposition 1,
with the exception of condition (vi), which was taken as:\medskip

Figure 10: $\ $(vi) $\ M=10^{3}$ and $N=1$.\medskip

Figure 11: $\ $(vi) $\ M=10^{3}$ and $N=10^{2}$.\medskip

Figure 12: $\ $(vi) $\ M=10^{3}$ and $N=10^{3}$.\medskip

 \begin{figure}[t]

 \centerline{\scalebox{0.7}{\includegraphics{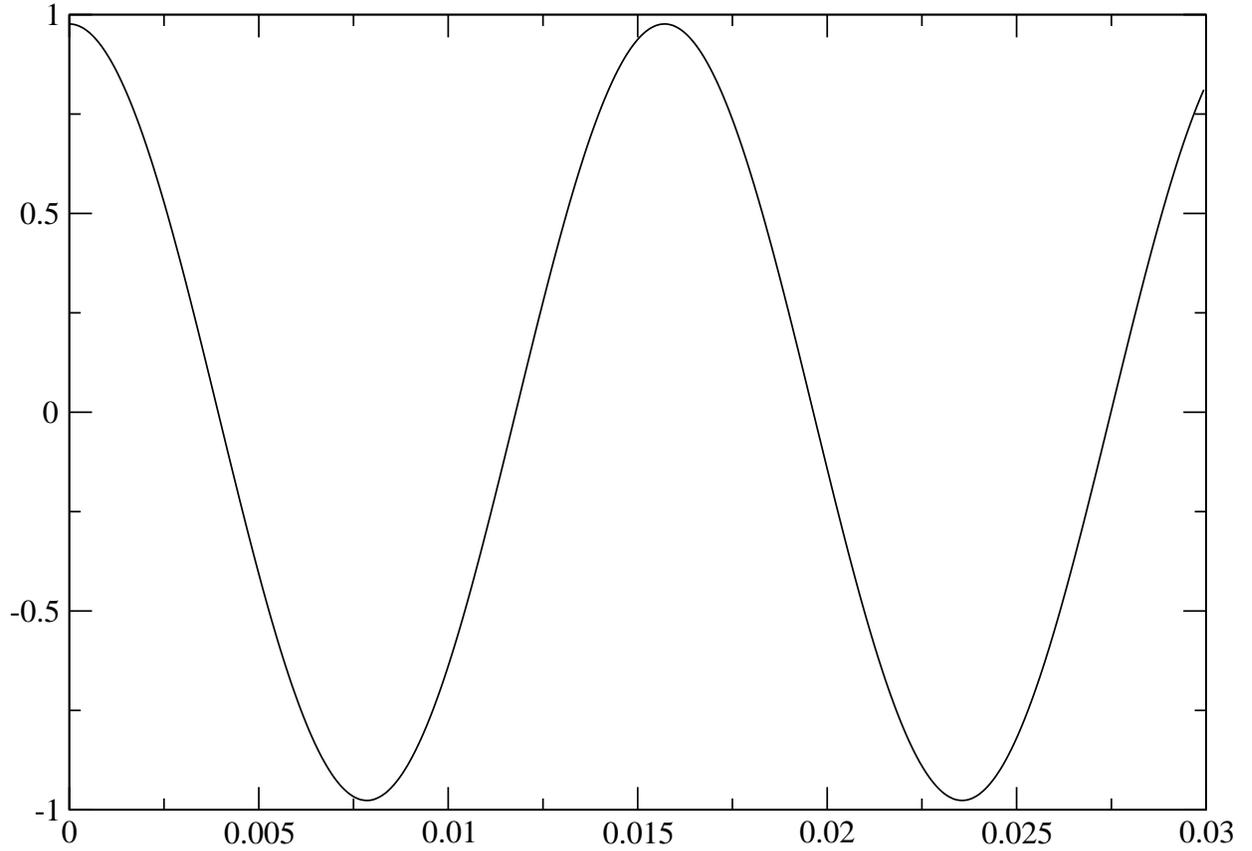}}}
\caption{Figure 10:
Evolution of $\Sigma ^{nd}(t)$ for $M=10^{3}$ and $N=1$, with $%
t_{0}=3.10^{-2}s$.}
 \label{f1}\vspace*{0.cm}
\end{figure}

 \begin{figure}[t]

 \centerline{\scalebox{0.7}{\includegraphics{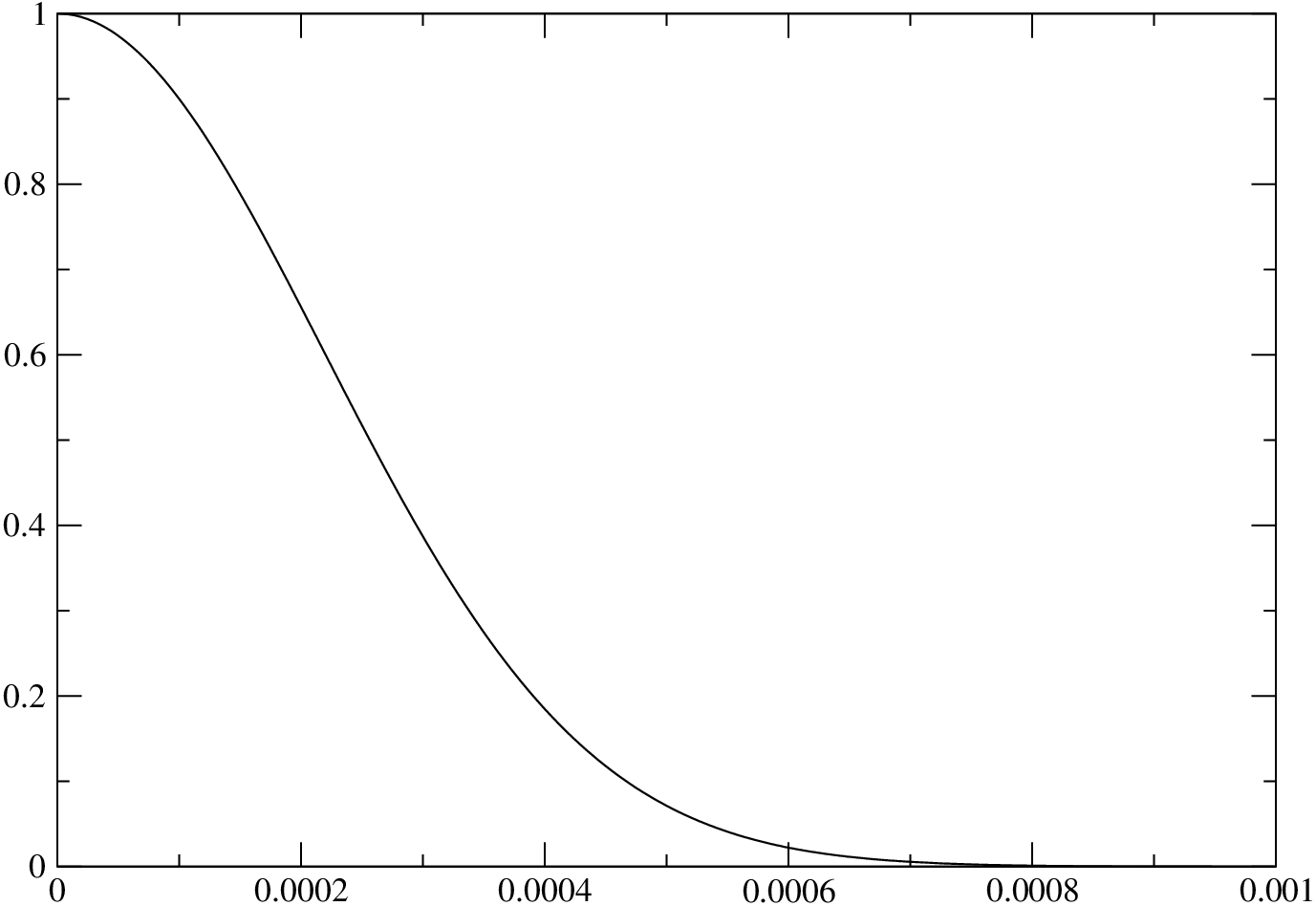}}}
\caption{Figure 11: Evolution of 
$\Sigma ^{nd}(t)$ for $M=10^{3}$ and $N=10^{2}$, with $t_{0}=1.10^{-3}s$.}
 \label{f1}\vspace*{0.cm}
\end{figure}

 \begin{figure}[t]

 \centerline{\scalebox{0.7}{\includegraphics{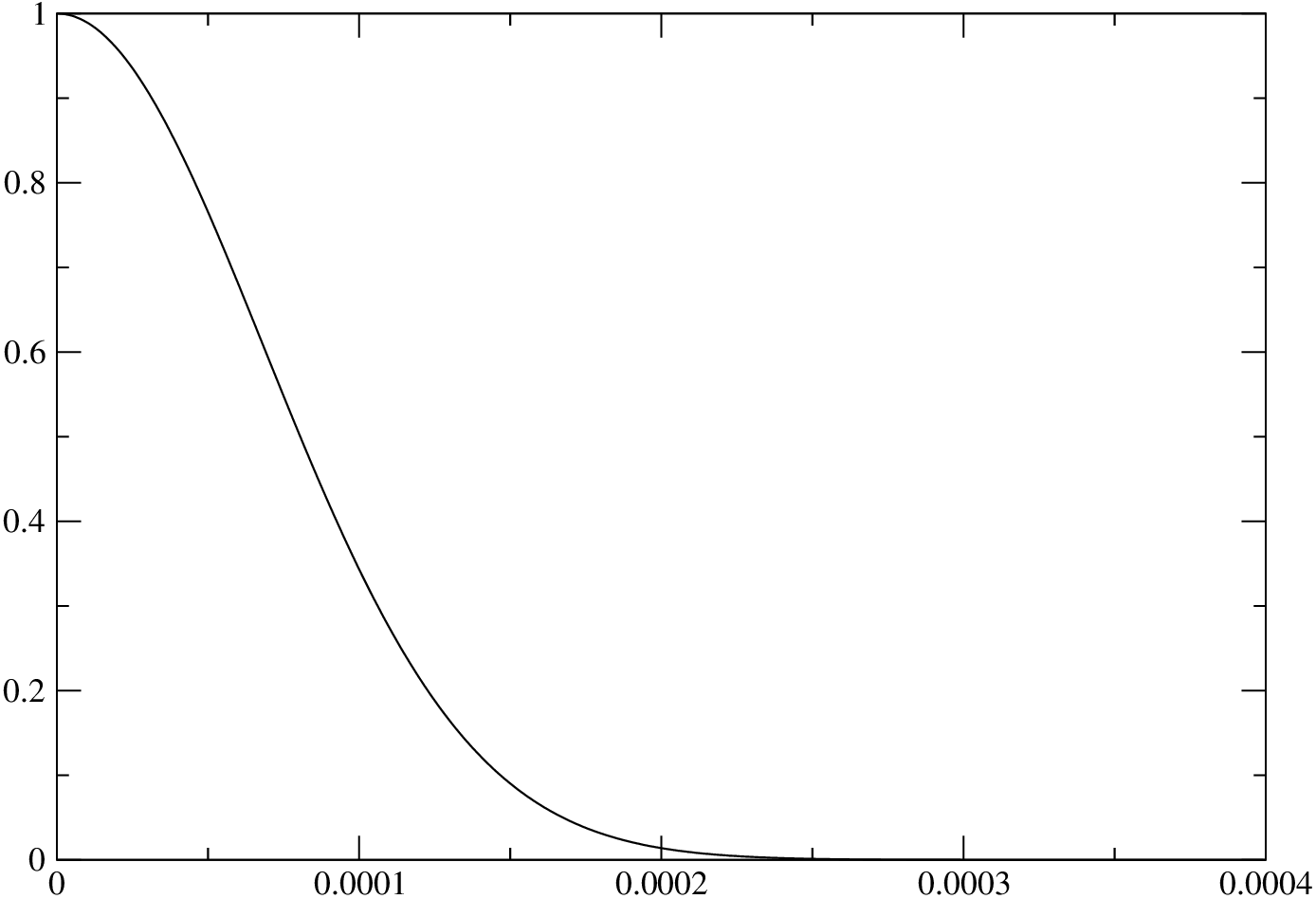}}}
\caption{Figure 12: Evolution of 
$\Sigma ^{nd}(t)$ for $M=10^{3}$ and $N=10^{3}$, with $t_{0}=4.10^{-4}s$.}
 \label{f1}\vspace*{0.cm}
\end{figure}

\textbf{Summarizing results}

As we have seen, in this decomposition of the whole closed system, the open
system $S=A_{M}$ decoheres when $N\gg 1$, independently of the value of $M$.
But the particle $A_{M}$ was selected as $S$ only for computation
simplicity: the same argument can be developed for any particle $A_{i}$ of $%
A $. Then, when $N\gg 1$ and independently of the value of $M$, any particle 
$A_{i}$ decoheres in interaction with its environment $E$ of $N+M-1$
particles.

On the other hand, as in Decomposition 1, here the symmetry of the whole
system $U$ allows us to draw analogous conclusions when the system $S$ is
one of the particles of $B$, say, $B_{N}$: $S=B_{N}$ decoheres when $M\gg 1$%
, independently of the value of $N$. And, on the basis of the same
considerations as above, when $M\gg 1$ and independently of the value of $N$%
, any particle $B_{i}$ decoheres in interaction with its environment $E$ of $%
N+M-1$ particles.

\section{Discussion}

\subsection{Analyzing results}

According to the usual reading of the EID approach, the decoherence of an
open system is induced by its interaction with a large environment. Such an
interaction is what leads to the dissipation of energy from the open system $%
S$ to the environment $E$. So, the orthodox view suggests a picture of
decoherence where an energy flow from the open system $S$ to the environment 
$E$ washes out the original coherence and allows the classicality of $S$ to
emerge. According to this picture, in the original spin-bath model a spin-$%
1/2$ particle $S=P$ decoheres when immersed in a large bath of spin-$1/2$
particles $E=\cup _{i}P_{i}$: $P$ dissipates its energy into $E$ and may
acquire a classical nature. However, this reading has to face the
\textquotedblleft looming big\textquotedblright\ problem of defining the
open systems involved in decoherence, since it does not provide a criterion
to identify the open system $S$ and its environment $E$. Now we will discuss
the results obtained in the generalized spin-bath model, in order to see how
they may contribute to the clarification of the problem.\medskip

a) As we have seen, in our generalized model, where $U=A\cup B$, with $A$ of 
$M$ particles $A_{i}$ and $B$ of $N$ particles $B_{i}$, (i) when $M\gg N$ or 
$M\simeq N$, the subsystem $A$ does not decohere (Decomposition 1 of Section
VII), but (ii) the particles $A_{i}$, considered independently, decohere
when $N\gg 1$ (Decomposition 2 of Section VIII). This means that there are
physically meaningful situations, given by $M\gg N\gg 1$ or $M\simeq N\gg 1$%
, where all the $A_{i}$ decohere although $A$ does not decohere. In other
words, in spite of the fact that certain particles decohere and may behave
classically, the subsystem composed by all of them retains its quantum
nature. This seemingly paradoxical conclusion sounds even more strange when
the situation is conceived in terms of energy dissipation. In spite of the
fact that all the $A_{i}$ dissipate their energy into the environment
(mainly into the subsystem $B$ due to the interaction among each $A_{i}$ and
all the $B_{i}$), the composite system $A=\cup _{i}A_{i}$ (which should
dissipate the energy of all the $A_{i}$) does not decohere.\medskip

b) We have also seen that, by symmetry, all the particles $B_{i}$,
considered independently, also decohere when $M\gg 1$. Then, when $M\gg N\gg
1$ or $M\simeq N\gg 1$, the requirement $M\gg 1$ holds and we can conclude
that not only all the $A_{i}$, but also all the $B_{i}$ decohere. So, all
the particles of the closed system $U=$ $\left( \cup _{i}A_{i}\right) \cup
\left( \cup _{j}B_{j}\right) $ may become classical when considered
independently, although the whole system $U$ certainly does not decohere
and, therefore, retains its quantum character. Again, the explanation of
this result is even more difficult when it is conceived in terms of the
energy dissipated from the system that decoheres to its environment, since
we are committed to decide which particles give and which receive the
dissipated energy.\medskip

These difficulties are further consequences of the \textquotedblleft looming
big\textquotedblright\ problem of defining the open systems involved in
decoherence. The irony of this story is that such a problem is the
consequence of what has been considered to be the main advantage of the
decoherence program: its open-system perspective. According to this
perspective, particles interacting with other particles by exchanging energy
are well-defined open systems, and the collections of those particles are
open systems too. \ So, the problem is to decide which one of all these open
systems is the system that decoheres or, in other words, where to place the
cut between the system $S$ and its environment $E$.

The open-system approach not only leads to the \textquotedblleft looming
big\textquotedblright\ problem, but also disregards the well-known holism of
quantum mechanics: a quantum system in not the mere collection of its parts
and its interactions. In order to retain its holistic nature, a quantum
system has to be considered as a whole: the open \textquotedblleft
subsystems\textquotedblright\ are only partial descriptions of the whole
closed system, given by the selection of particular subspaces of relevant
observables. On the basis of this closed-system perspective, we can develop
a different conceptual viewpoint for understanding decoherence, which
dissolves the problems of the orthodox open-system view.

\subsection{A different conceptual viewpoint}

As we have seen, a TPS expresses the decomposition of the closed system $U$,
represented in the Hilbert space $\mathcal{H}=\mathcal{H}_{A}\otimes 
\mathcal{H}_{B}$, into two open systems $S_{A}$ and $S_{B}$, represented in $%
\mathcal{H}_{A}$ and $\mathcal{H}_{B}$ respectively. Such a decomposition
amounts to the split of the whole space $\mathcal{O}=\mathcal{H}\otimes 
\mathcal{H}$ of the observables of $U$ into the subspaces $\mathcal{O}_{A}=%
\mathcal{H}_{A}\otimes \mathcal{H}_{A}$ and $\mathcal{O}_{B}=\mathcal{H}%
_{B}\otimes \mathcal{H}_{B}$ such that $\mathcal{O}=\mathcal{O}_{A}\otimes 
\mathcal{O}_{B}$. \ In particular, the total Hamiltonian of $U$, $H\in 
\mathcal{O}$, can be expressed as $H=H_{A}\otimes I_{B}+I_{A}\otimes
H_{B}+H_{AB}$, where $H_{A}\in \mathcal{O}_{A}$ is the Hamiltonian of $S_{A}$%
, $H_{B}\in \mathcal{O}_{B}$ is the Hamiltonian of $S_{B}$, and $H_{AB}\in 
\mathcal{O}$ is the interaction Hamiltonian, representing the interaction
between the systems $S_{A}$ and $S_{B}$.

As stressed in papers \cite{Sujeeva-1} and \cite{Sujeeva-2}, in general a
quantum system $U$ admits a variety of TPSs, that is, a variety of different
decompositions into $S_{A}$ and $S_{B}$, each one defined by the space of
observables $\mathcal{O}_{A}$ of $S_{A}$ and $\mathcal{O}_{B}$ of $S_{B}$.
Among all the possible decompositions of $U$, there is a particular TPS that
remains \textit{dynamically invariant}. This is the case when the
interaction Hamiltonian $H_{AB}=0$: there is no interaction between $S_{A}$
and $S_{B}$ and, then, 
\begin{equation}
\left[ H_{A}\otimes I_{B},I_{A}\otimes H_{B}\right] =0\quad \Longrightarrow
\quad \exp \left( -iHt\right) =\exp \left( -iH_{A}t\right) \exp \left(
-iH_{B}t\right)  \label{7.1}
\end{equation}%
Therefore, 
\begin{eqnarray}
\rho _{A}(t) &=&Tr_{(B)}\rho (t)=e^{iH_{A}t}\left( Tr_{(B)}\rho _{0}\right)
\,e^{-iH_{A}t}=e^{iH_{A}t}\rho _{A0}\,e^{-iH_{A}t}  \label{7.3} \\
\rho _{B}(t) &=&Tr_{(A)}\rho (t)=e^{iH_{B}t}\left( Tr_{(A)}\rho _{0}\right)
\,e^{-iH_{B}t}=e^{iH_{B}t}\rho _{B0}\,e^{-iH_{B}t}  \label{7.4}
\end{eqnarray}%
This means that, even if the initial state $\rho _{0}$ of $U$ is an
entangled state with respect to the TPS $\mathcal{H}=\mathcal{H}_{A}\otimes 
\mathcal{H}_{B}$, the subsystems $S_{A}$ and $S_{B}$ are \textit{dynamically
independent}: each one of them evolves \textit{unitarily} under the action
of its own Hamiltonian. As a consequence, the subsystems $S_{A}$ and $S_{B}$
resulting from this particular, dynamically invariant TPS do not decohere.

Once we have excluded the dynamically invariant TPS of $U$, all the
remaining TPSs define interacting subsystems $S_{A}$ and $S_{B}$, such that $%
H_{AB}\neq 0$. As a result of the interaction, $S_{A}$ and $S_{B}$ evolve
non-unitarily and, then, depending on the particular interaction between
them, they may decohere. But the point to stress here is that \textit{there
is no privileged non-dynamically invariant decomposition of }$U$: each
partition of the closed system into $S_{A}$ and $S_{B}$ is just a way of
selecting the spaces of observables $\mathcal{O}_{A}$ and $\mathcal{O}_{B}$.

When we adopt a closed-system perspective by means of the concept of TPS, it
turns out to be clear that, in decoherence, there is no essential criterion
for identifying the \textquotedblleft open system\textquotedblright\ and its
\textquotedblleft environment\textquotedblright . Given the closed system $U$%
, that identification requires two steps: (i) to select a TPS $\mathcal{H}=%
\mathcal{H}_{A}\otimes \mathcal{H}_{B}$, such that $U=S_{A}\cup S_{B}$, and
(ii) to decide that one of the systems resulting from the decomposition, say 
$S_{A}$, is the open system $S$, and the other, $S_{B}$, is the environment $%
E$. Since the TPS is defined by the spaces of observables $\mathcal{O}_{A}$
and $\mathcal{O}_{B}$, the decomposition of $U$ is just the adoption of a
descriptive perspective: the identification of $S$ and $E$ amounts to the
selection of the observables relevant in each situation. But since the split
can be performed in many ways, with no privileged or essential
decomposition, there is no need of an unequivocal criterion for deciding
where to place the cut between \textquotedblleft the\textquotedblright\
system and \textquotedblleft the\textquotedblright\ environment. Decoherence
is not a yes-or-not process, but a phenomenon \textit{relative} to the
chosen decomposition of the whole closed quantum system. When viewed from
this closed-system perspective, Zurek's \textquotedblleft looming big
problem\textquotedblright\ does not constitute a real threat to the
decoherence program: the supposed challenge dissolves once the relative
nature of decoherence is taken into account.

>From this perspective, quantum mechanics is a theory whose dynamical
postulate refers to closed systems: the time-behavior of the parts resulting
from different partitions of the closed system has to be inferred from that
postulate. Since the total Hamiltonian rules the dynamical evolution of the
closed system, then the time-behavior of its open subsystems depends on the
form in which the Hamiltonian is decomposed in each particular partition.
This means that decoherence cannot be simply described as the result of an
interaction through which a small open system $-$typically, a particle$-$
dissipates its energy into a large environmental bath. As we have seen in
the generalized spin-bath model, this picture of decoherence leads to
perplexities: the relationships between the whole closed system and its open
subsystems is subtler than that picture suggests (in a future paper we will
study those relationships from a theoretical viewpoint in order to draw some
general conclusions regarding decoherence). Therefore, the decomposition of
the total Hamiltonian has to be studied in detail in each particular case,
in order to know whether the system of interest resulting from the partition
of the whole closed system decoheres or not under the action of its
self-Hamiltonian and the interaction Hamiltonian.

\section{Conclusions}

The aim of this paper has been to argue that decoherence can be viewed from
a closed-system perspective, which improves the understanding of the
phenomenon. For this purpose, we have analyzed the simple spin-bath model by
studying the time-behavior of the expectation values of relevant observables
belonging to different sets. Then, we have generalized the original model in
order to see how decoherence depends on the way in which the relevant
observables are selected.

On the basis of the analysis of the two models from a closed-system
perspective, we have drawn the following conclusions:

\begin{enumerate}
\item[(i)] Decoherence is a phenomenon relative to which degrees of freedom
of the whole closed system are considered relevant and which are disregarded
in each case.

\item[(ii)] The explanation of decoherence requires the detailed study of
the interaction Hamiltonian resulting from the selected partition of the
whole closed system.

\item[(ii)] Although it is usually claimed that EID is a dissipative
approach to decoherence, the simple account of decoherence in terms of
energy dissipation from the open system to its environment is misguided, to
the extent that there are situations where all the particles of a closed
system decohere when considered independently.

\item[(iv)] Since there is no privileged or essential decomposition of the
closed system, there is no need of an unequivocal criterion for identifying
the systems involved in decoherence. Therefore, the \textquotedblleft
looming big problem\textquotedblright , which, according to Zurek, poses a
serious threat to the whole decoherence program, looses its strength in the
light of the relative nature of decoherence.
\end{enumerate}

\end{document}